\documentclass[journal]{IEEEtran}
\usepackage{cite}
\usepackage{amsmath,amssymb,amsfonts}
\usepackage{algorithmic}
\usepackage{graphicx}
\usepackage{textcomp}
\usepackage{harpoon}
\usepackage{amsmath}
\usepackage{mathrsfs}
\usepackage{amssymb}
\usepackage{epstopdf}
\usepackage{subfig}
\usepackage{lmodern}
\usepackage{bm}
\usepackage{sansmath}
\usepackage{booktabs}
\usepackage{colortbl}
\usepackage{esint}
\usepackage{stfloats}
\usepackage{lipsum}
\usepackage{multicol}
\usepackage{multirow}
\usepackage{tabularx}
\usepackage{color}
\usepackage{cases}

\ifCLASSINFOpdf

\else

\fi

\usepackage[absolute,showboxes]{textpos}

\TPMargin{3pt}
\newcommand{\copyrightstatement}{
	\begin{textblock}{0.84}(0.08,0.95) 
		\noindent
		\footnotesize
		\copyright 2023 IEEE. Personal use of this material is permitted. Permission from IEEE must be obtained for all other uses, in any current or future media, including reprinting/republishing this material for advertising or promotional purposes, creating new collective works, for resale or redistribution to servers or lists, or reuse of any copyrighted component of this work in other works. DOI: 10.1109/TWC.2023.3325042.
	\end{textblock}
}

\begin{document}
	
\copyrightstatement
\title{Decomposed and Distributed Directional Modulation for Secure Wireless Communication}

\author{
	Bin Qiu, \textit{Member, IEEE}, Wenchi Cheng, \textit{Senior Member, IEEE}, \\ and Wei Zhang, \textit{Fellow, IEEE}
\thanks{
}
\thanks{
	This work was supported in part by the National Key R\&D Program of China under Grant 2021YFC3002102, in part by the Key R\&D Plan of Shaanxi Province under Grant 2022ZDLGY05-09, in part by the Key Area R\&D Program of Guangdong Province under Grant 2020B0101110003, in part by the Fundamental Research Funds for the Central Universities under Grant XJS220105, in part by the Project funded by China Postdoctoral Science Foundation under Grant 2022M712491, and in part by the Natural Science Basic Research Program of Shaanxi under Grant 2023-JC-QN-0715. (Corresponding author: Wenchi Cheng.)}
\thanks{
	Bin Qiu and Wenchi Cheng are with the State Key Laboratory of Integrated Services Networks, Xidian University, Xian
	710071, China (e-mail: qiubin@xidian.edu.cn; wccheng@xidian.edu.cn).	
	}
\thanks{
	Wei Zhang is with the School of Electrical Engineering and Telecommunications, University of New South Wales, Sydney, NSW 2052, Australia (e-mail: w.zhang@unsw.edu.au).
}

}

\maketitle

\begin{abstract}
Directional modulation and artificial noise (AN)-based methods have been widely employed to achieve physical-layer security (PLS). However, these approaches can only achieve angle-dependent secure transmission. This paper presents an AN-aided decomposed and distributed directional modulation (D3M) scheme for secure wireless communications, which takes advantage of the spatial signatures to achieve an extra range-dimension security apart from the angles. Leveraging decomposed and distributed structure, each of modulated signal is represented by mutually orthogonal in-phase and quadrature branches, which are transmitted by two distributed transmitters to enhance PLS. In particular, we first aim to minimize transmit message power by integrated design of the transmit beamformers, subject to prescribed received signal-to-noise ratio (SNR) for the legitimate user (LU) and no inter-branch interference. This guarantees reliable and accurate transmission for the LU with the minimum transmit message power. Considering the leakage power on the sidelobes, AN is superimposed on the messages to try to mask the confidential information transmission. Simulation results demonstrate the security enhancement of our proposed D3M system.

\end{abstract}


\IEEEpeerreviewmaketitle

\section{Introduction}
\IEEEPARstart{A}{long} with the countless applications of wireless communication, its security is particularly important but presents many challenges, as the broadcast nature of wireless medium makes the confidential messages exposed to surrounding adversaries and vulnerable to malicious intercepting. Upon traditional upper-layer data encryption, physical-layer security (PLS) \cite{Principles_Mukherjee,Physical_Liu}, as a security-enhancing technique, has attracted great interests in the past decade. In PLS of wireless communications, the key idea is to utilize the intrinsic nature of wireless medium to encrypt data transmission from transmitter to intended users, while defending the confidential messages from wiretapping \cite{Survey_Wu}.

Phased-array transmission based directional modulation (DM) technique is to employ an array of antennas at the transmitter, in which the relative phases of the respective signals feeding the antennas are modified. In such a way, the effective radiation pattern of the array is reinforced along a predefined direction in the free space and debilitated in other directions. Several related approaches have been used in PLS of wireless communications. The authors in \cite{Directional_Daly} enhanced PLS by optimizing a set of phase shifters across array antennas to generate a standard constellation along the given direction and distort the reception in the undesired directions. The wiretap channel model was first introduced by Wyner \cite{Wireless_Bloch}, where proved that the perfect information-theoretic security can be achieved by using PLS techniques. In this model, the concept of secrecy rate was introduced, also referred to as secrecy capacity, to assess the secrecy performance. Inspired by this secrecy metric, the authors of \cite{Opportunistic_Abbas,Joint_Nguyen,Artificial_Wang} adopted a secrecy rate maximization criterion to enhance PLS. However, the calculation of secrecy rate needs to acquire the precise location knowledge of the eavesdroppers (Eves). Unfortunately, passive Eves usually keep radio silence to avoid exposure. Therefore, it is not feasible to get any information on the passive Eves in practice. A closely related work on phased-array DM structure was studied in \cite{Phased_Zhang}, where the phase shift keying (PSK) modulation was used for secure mmWave wireless communications via polygon construction in the complex plane. The aid of polygon construction synthesized exact/relaxed phases at legitimate user (LU) and produced random disruption in other directions. 

Another efficient way for guaranteeing secure communication is the embedding of artificial noise (AN) \cite{Frequency_Qiu}, also called as jamming noise, at the transmitter, which is transmitted simultaneously with messages to interfere with Eves. The emission of AN can seriously degrade the recovery of messages in the undesired directions. In  \cite{Artificial_Xie,Multi_Chen}, the authors considered a passive Eve case where AN is designed to null out the interference to LU. The authors of \cite{Artificial_Zhao} presented an AN-assisted interference alignment scheme with wireless power transfer. The total transmit power of AN is maximized by jointly optimizing the information power and the coefficient of power splitting. To withstand the imperfect direction knowledge, several robust synthesis schemes of phased-array transmission were explored as an emerging subject in different works \cite{Robust_Shu,Robust_Qiu,Robust_Ng}. Furthermore, the authors in \cite{Directional_Shu,Hybrid_Wang} revealed a new array DM system that a hybrid phased multiple-input multiple-output (MIMO) structure was developed to improve the secrecy capacity for future fifth generation (5G) cellular systems. 

Considering a practical scenario, Eves may exactly locate along an identical direction as the LU to wiretap the confidential messages. It fails to provide PLS by using the conventional DM approach based on phased arrays in this case. This is due to the theoretical limitation that the beampattern of phased-array transmission can only achieve angle-dependent secure transmission. Consequently, it is of interest to seek to new secure transmission methods that can promote to an extra layer of security by introducing additional flexibility. The frequency diverse array (FDA) adds a small frequency offset across the array antennas, which results in exhibiting an extra range-dimension dependence secure transmission \cite{WFRFT_Cheng,Fixed_Hong}. In particular, the authors in \cite{Physical_Lin} jointly designed the frequency offsets and the beamforming vector of the FDA transmitter to enhance PLS for proximal LU and Eve. A random subcarrier selection scheme based on orthogonal frequency division multiplexing (OFDM) was employed in \cite{Secure_Shu} to achieve the precise wireless communication with a low-complexity structure. However, the radiation pattern of the FDA is range-time-coupled. This characteristic causes the power peak in a specific range to change over time, in addition to the very complicated synchronization for the LU \cite{Index_Qiu}. What's more, the authors in \cite{Time_Ding} pointed out that it is impossible for the FDA enhanced DM system to realize wireless security in range dimension when Eves sample the signals at the different time instant as LU. In addition, most of the prior work on beamforming design for secure transmission assumed that the information of the Eve is perfectly known at the transmitter. However, those assumption is not practical, particularly when Eves are passive devices.

To overcome these limitations of the previous works, in this paper we propose an AN-aided decomposed and distributed directional modulation (D3M) secure transmission scheme in the presence of passive Eves, where the information of Eves is assumed to be unknown at transmitter. More specifically, the decomposed in-phase and quadrature branches are transmitted by two distributed transmitters. In this way, the signals emitted by one transmitter do not carry full information directly. That is, the effective receiving zone only happens in the intersection of the two mainlobes from the distributed transmitters. Therefore, the proposed scheme is capable of providing angle-range-dependent “point” secure transmission, whose security overcomes the “line” secure transmission of the conventional phased array transmission. By leveraging distributed phased-array beamforming, we seek to minimize transmit message power by designing the transmit beamforming vectors to reduce the potential of message power leakage, subject to satisfying a prescribed received signal-to-noise ratio (SNR) constraint of each branch to provide satisfactory communication performance for LU. Unlike existing AN design approaches \cite{Robust_Shu} that the design of AN is only to force AN into the null space of LUs, we aim to minimizes transmit AN power subject to target signal-to-interference-plus-noise ratio (SINR) constraints for the undesired directions.

The rest of this paper is organized as follows. Section II introduces the system model of the AN-aided D3M secure transmission scheme. We formulate the optimization problem; then, we develop algorithms to solve the problem and extend our proposed scheme to the case of multi-LU system in Section III. We analyze the secrecy performance in Section IV. Simulation results are presented in Section V. Finally, Section VI contains our concluding remarks.

\textit{Notations}: Matrices, vectors, and scalars are denoted by bold upper-case, bold lower-case, and lower case letters to denote , respectively. $\mathbb{E}\{\cdot\}$ and tr$(\cdot)$ stand for the expectation and trace operations, respectively. vec$(\cdot)$ stacks columns of matrix into a single column vector. diag$(\cdot)$ constructs a diagonal matrix with the elements of its argument on the diagonal. $\left\|  \cdot  \right\|_2$ and $\left|  \cdot  \right|$ represent the $\ell_2$-norm and modulus, respectively. Superscripts $( \cdot )^ {-1} $, $( \cdot )^T$, and $( \cdot )^H$ denote the inverse, transpose, and Hermitian, respectively. Re$\{\cdot\}$ and Im$\{\cdot\}$ extract the real part and the imaginary part of its argument. We denote by Matrices ${{\mathbf{I}}_N}$ and ${{\boldsymbol{0}}_{N \times M}}$ the identity matrix with $N \times N$ and zeros matrix with $N \times M$, respectively. Use $\mathbb{R}$ and $\mathbb{C}$ to indicate real and complex number field, respectively. 
\section{System Sketch and Signal Model}
\begin{figure}
	\centering
	\includegraphics[width=0.9\columnwidth]{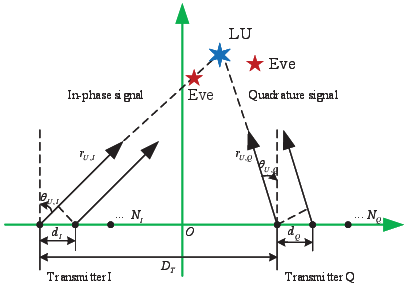}
	\caption{System model of the proposed AN-aided D3M secure transmission scheme.}
	\label{fig1}
\end{figure}
Let us consider a wireless communication system, as shown in Fig. \ref{fig1}, that consists of two transmitters with physical separation,  transmitter I and Q, equipped with an ${N_I}$- and ${N_Q}$- element array, respectively, and one single-antenna LU. There exist one or several single-antenna passive Eve whose information is unavailable at the transmitters. The two transmitters forward mutually orthogonal in-phase and quadrature branches decomposed by $M$-PSK ($M>2$) modulation signal to the LU in the presence of passive Eves wiretapping the confidential messages. For simplicity, we employ a uniform linear array (ULA) consisted of isotropic antennas for transmitters, and the results can be easily extended to multidimensional periodic arrays. Without loss of generality, the first element of the ULA is defined as phase reference. Let $d_I$ and $d_Q$ denote the ULA's adjacent elements spacing of the transmitter I and Q, respectively. We assume that the transmitters are symmetrical about the origin along the x-axis, and the distance is $D_T$ with $d_I \ll D_T$ and $d_Q \ll D_T$. For similarity, we ignore the very few multi-path components in high frequency transmission, and the far-field parallel wavefront and line-of-sight (LoS) assumptions can hold simultaneously due to the tiny array size. For an arbitrary receiver located at $(x,y)$ on the x-y plane, the superimposed electric field radiation of the distributed transmit array in the free-space path loss channel, denoted by $B({r_I}, {\theta_I}, {r_Q}, {\theta_Q})$, is given by \cite{Optimum_Van}
\begin{align}
	&B({r_I}, {\theta_I}, {r_Q}, {\theta_Q}, t)\notag\\&
	= {\rho _I}{\mathop{\rm Re}\nolimits} \{ \sum\limits_{n = 1}^{{N_I}} {{w_{I,n}}{e^{j2\pi {f_c}[t - \frac{{{r_I} - \left( {n - 1} \right){d_I}\sin {\theta _I}}}{c}]}}} \}\notag \\&  \kern 6pt
	+{\rho _Q}{\mathop{\rm Im}\nolimits} \{ \sum\limits_{m = 1}^{{N_Q}} {{w_{Q,m}}{e^{j2\pi {f_c}[t - \frac{{{r_Q} - \left( {m - 1} \right){d_Q}\sin {\theta _Q}}}{c}]}}} \} \notag \\&
	= {\rho _I}{\rm Re} \{{e^{2\pi {f_c}(t - \frac{{{r_I}}}{c})}} \sum\limits_{n = 1}^{{N_I}} {{w_{I,n}}{e^{\frac{{j2\pi {f_c}(n - 1){d_I}\sin {\theta _I}}}{c}}}} \}\notag \\& \kern 6pt
	+ {\rho _Q}{\rm Im} \{{e^{2\pi {f_c}(t - \frac{{{r_Q}}}{c})}} \sum\limits_{m = 1}^{{N_Q}} {{w_{Q,m}}{e^{\frac{{j2\pi {f_c}(m - 1){d_Q}\sin {\theta _Q}}}{c}}}} \} ,
	\label{eq1}\end{align}
where $f_c$ denotes the carrier frequency, $c$ represents the speed of light, $w_{I,n}$ and $w_{Q,m}$ denote the beamformer at transmitter I and Q, respectively, $\rho_I$ and $\rho_Q$ stand for the path loss factor related to the distance, $({r_I},{\theta_I})$ and $({r_Q},{\theta_Q})$ respectively denote the coordinate. $\footnote{{According to the basic geometric manipulations presented in Fig. \ref{fig1}, the location (range, angle) can be obtained by using coordinate transformation as ${r_I}=\sqrt {{{(x + D_T/2)}^2} + {y^2}}$, ${\theta_I}=\arctan [(x + D_T/2)/y]$, ${r_Q}=\sqrt {{{(x - D_T/2)}^2} + {y^2}}$, and ${\theta_Q}=\arctan [(x - D_T/2)/y]$.}}$ Let us define the channel vector from transmitter I and Q to receiver as
\begin{eqnarray}
	\begin{aligned}[b]
		\left\{ {\begin{array}{*{20}{l}}
				{{{\bf{h}}_I}({r_I},{\theta_I}) \buildrel \Delta \over = {\rho _I}{[{e^{{\Phi _{{_I}}}(1)}},{e^{{\Phi _{{_I}}}(2)}},...,{e^{{\Phi _{{ _I}}}({N_I})}} ]^H},}\\
				{{{\bf{h}}_Q}({r_Q},{\theta_Q}) \buildrel \Delta \over = {\rho _Q}{[{e^{{\Phi _{{_Q}}}(1)}},{e^{{\Phi _{{_Q}}}(2)}},...,{e^{{\Phi _{{_Q}}}({N_Q})}} ]^H},}
		\end{array}} \right.
	\end{aligned}
	\label{eq2}\end{eqnarray}
where ${\Phi _I}(n) = \frac{{ j 2\pi  {f_c}( {n - 1} ){d_I}\sin {\theta _I}}}{c},n \in {\mathcal{{\cal N}}_I}$, ${\Phi _Q}(m) = \frac{{ j 2\pi {f_c}( {m - 1} ){d_Q}\sin {\theta _Q}}}{c},m \in {\mathcal{{\cal N}}_Q}$, ${\mathcal{{\cal N}}_I} \buildrel \Delta \over = \{ 1,2,...,{N_I}\}$, and ${\mathcal{{\cal N}}_Q}\buildrel \Delta \over = \{ 1,2,...,{N_Q}\}$. To combat the effect of path loss on the received power, the path loss factor is absorbed in the channel vector, yielding the standard received constellation diagram for LU.

\textsl{Remark 1:} For the case of non-light-of-sight (NLoS) or fading channel, distributed MIMO precoding method is a way to achieve secure transformation. Following the idea of our decomposed and distributed transformation,  the decomposed orthogonal branches (such as OFDM) are transmitted by distributed MIMO transmitters. The design of distributed MIMO precoding is also to guarantee efficient reception of each link.
\begin{figure}
	\centering
	\includegraphics[width=1\columnwidth]{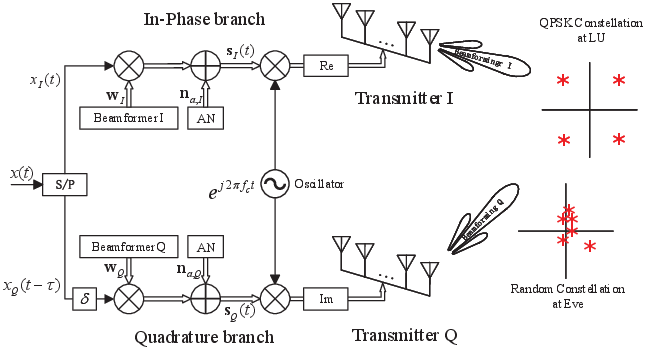}
	\caption{Generic architecture of the D3M transmitters, including the optimal beamforming vector and AN.}
	\label{fig2}
\end{figure}

In our proposed scheme, an $M$-PSK constellation symbol is divided into in-phase and quadrature modulation symbol components at the beginning of the transmission process, and then modulate them to mutually orthogonal in-phase and quadrature branches, each of which is emitted by a transmitter. Based on this idea, as shown in Fig. \ref{fig2}, the instantaneous AN-aided baseband transmit signals at transmitter I and Q, denoted by ${{\bf{s}}_I}(t)$ and ${{\bf{s}}_Q}(t)$, are given by
\begin{eqnarray}
	\begin{aligned}[b]
\left\{ {\begin{array}{*{20}{l}}
		{{{\bf{s}}_I}(t) = {{\bf{w}}_I}{x_I}(t) + {{\bf{n}}_{A,I}},}&\\
		{{{\bf{s}}_Q}(t) = \delta  \cdot {{\bf{w}}_Q}{x_Q}(t - \tau ) + {{\bf{n}}_{A,Q}},}
\end{array}} \right.
\end{aligned}
\label{eq3}\end{eqnarray} 
where ${{\bf{w}}_I}  \in \mathbb{C}^{{N_I} \times 1}$ is the beamforming vector at transmitter I with ${{\bf{w}}_I} = {[{w_{I,1}},...,{w_{I,{N_I}}}]^T}$, ${{\bf{w}}_Q} \in \mathbb{C}^{{N_Q} \times 1}$ is the beamforming vector at transmitter Q with ${{\bf{w}}_Q} = {[{w_{Q,1}},...,{w_{Q,{N_Q}}}]^T}$, ${x_I}(t) \in \mathbb{R}$ and ${x_Q}(t)\in \mathbb{R}$ are the in-phase and quadrature symbol components after serial-to-parallel (S/P) conversion of the input bipolar message sequence ${x}(t)\in \mathbb{C}$ \cite[Ch. 5]{Wireless_Goldsmith}, ${x}(t)={x_I}(t)+j{x_Q}(t)$, with $\mathbb{E}\{{\left| {x}(t) \right|^2}\} = 1$, $\delta $ is the phase compensation factor, which is employed at transmitter Q to guarantee the orthogonality of the carrier, $\tau$ is latency factor for symbol alignment, and ${{\mathbf{n}}_{A,I}}$ and ${{\mathbf{n}}_{A,Q}}$ are AN emitted by transmitter I and Q, respectively. 

\textsl{Remark 2:}  Due to the coordinated operations for the distributed transmitters, we consider the communication system that consists of one communication control unit and two distributed arrays, which are physically separated by a distance, or two independent distributed transmit systems that each transmit array is controlled by one control unit, and they are cooperative by sharing information over wireless  \cite{Distributed_Ellison} or fiber communications.

At the beginning of each scheduling slot, LU sends handshaking/beacon signals to the transmitters to report their status (location information and service requirements). The information embedded in the handshaking signals facilitates the downlink packet transmission. Thus, the transmitter-to-LU fading gains can be reliably estimated at transmitters with negligible estimation errors. Define $(x_U,y_U)$  on the x-y plane as the coordinate of the LU corresponding location $({r_{U,I}},{\theta_{U,I}})$ relative to transmitter I, and $({r_{U,Q}},{\theta_{U,Q}})$ relative to transmitter Q. To simplify the notations, we use ${{\bf{h}}_{U,I}}$ and ${{\bf{h}}_{U,Q}}$ to denote the channel vectors from the transmitter I and Q to LU, i.e.,  ${{\bf{h}}_{U,I}} \buildrel \Delta \over = {{\bf{h}}_I}({r_{U,I}},{\theta _{U,I}})$ and ${{\bf{h}}_{U,Q}} \buildrel \Delta \over = {{\bf{h}}_Q}({r_{U,Q}},{\theta _{U,Q}})$. After passing through wireless channels, a mixed signal consisting of the in-phase and quadrature branches from transmitter I and Q arrives at LU. Therefore, the received signal at LU, denoted by ${y_U}(t)$, can be expressed as 
\begin{align}
{y_U}(t) = \underbrace {{\rm{Re}}\{ {g_{U,I}}{\bf{h}}_{U,I}^H{{\bf{s}}_I}\} }_{{\text{From transmitter I}}} + \underbrace {{\rm{Im}}\{ {g_{U,Q}}{\bf{h}}_{U,Q}^H{{\bf{s}}_Q}\} }_{{\text{From transmitter Q}}} + \underbrace {{n_U}}_{{\text{AWGN}}},
\label{eq4}\end{align}
where ${g_{U,I}} = {e^{j2\pi {f_c}(t - \frac{{{r_{U,I}}}}{c})}}$ and ${g_{U,Q}} = {e^{j2\pi {f_c}(t - \frac{{{r_{U,Q}}}}{c})}}$ are the carrier with the path delay, ${{{n}}_{U}}$ is the additive white Gaussian noise (AWGN) with zero-mean and variance $\sigma _U^2$, i.e., ${{{{n}}_{U}}}\sim{\mathcal{N}}({{{0}}},\sigma _U^2)$.
	
Similarly, let us denote by $({r_{E,I}},{\theta _{E,I}})$ and $({r_{E,Q}},{\theta _{E,Q}})$ the Eve's locations relative to the transmitters I and Q, respectively. We define ${{\bf{h}}_{E,I}}$ and ${{\bf{h}}_{E,Q}}$ as the channel vectors, i.e.,  ${{\bf{h}}_{E,I}} \buildrel \Delta \over = {{\bf{h}}_I}({r_{E,I}},{\theta _{E,I}})$ and ${{\bf{h}}_{E,Q}} \buildrel \Delta \over = {{\bf{h}}_E}({r_{E,Q}},{\theta _{E,Q}})$. Then, the received signal at Eve, denoted by ${y_E}(t)$, is given by
\begin{align}
	{y_E}(t) = \underbrace {{\rm{Re}}\{ {g_{E,I}}{\bf{h}}_{E,I}^H{{\bf{s}}_I}\} }_{{\text{From transmitter I}}} + \underbrace {{\rm{Im}}\{ {g_{E,Q}}{\bf{h}}_{E,Q}^H{{\bf{s}}_Q}\} }_{{\text{From transmitter Q}}} + \underbrace {{n_E}}_{{\text{AWGN}}},
\label{eq5} \end{align} 
where ${g_{E,I}} = {e^{j2\pi {f_c}(t - \frac{{{r_{E,I}}}}{c})}}$ and ${g_{E,Q}} = {e^{j2\pi {f_c}(t - \frac{{{r_{E,Q}}}}{c})}}$ are the carrier with the path delay, ${{{n}}_{E}}$ is the AWGN with zero-mean and variance $\sigma _E^2$, i.e., ${{{{n}}_{E}}}\sim{\mathcal{N}}({{{0}}},\sigma _E^2)$.	
	
\section{Proposed Secure Transmission Strategy}
To achieve PLS, it is critical to realize reliable communication with LU, but equally important to avoid wiretapping. Towards this end, we propose an effective scenario of designing the beamforming vectors and AN projection matrix.	

\subsection{Beamforming Vector Optimization}	
Our design is no attempt to place nulls in the locations of Eves due to the unavailable location information of passive Eves. To reduce the probability of interception, we seek to minimize the transmit message power such that less transmit message power can reduce the message leakage. Meanwhile, we need to strictly control the received SNR of per signal branch arrived at the LU-side, so that LU can decode the symbols, correctly. The basic idea of our proposed D3M transmission is that transmitter I emits the in-phase branch and transmitter Q emits the quadrature branch. Therefore, the design of the beamforming vectors of the distributed transmitters that minimizes the total transmit message power subject to meeting the constraint on received SNR requirement at LU can be formulated as$\footnote{The problem can also be designed as reverse transmission, i.e., transmitter I emits quadrature branch and transmitter Q emits in-phase branch. In this case, the real part is equal to 0 and the imaginary part exceeds the required SNR of the received signal.}$
\begin{subequations}
	\begin{equation}\textbf{P1:}  \kern 4pt \mathop {\min }\limits_{{{\bf{w}}_I},{{\bf{w}}_Q}} \kern 2pt \Vert {\bf{w}}_I \Vert_2^2 + \Vert {\bf{w}}_Q \Vert_2^2 \kern 46pt \tag{6a} \label{eq8a*}\end{equation}
	\begin{equation} {\rm{s.t.}} \kern 2pt \text{Re}\{{\bf{h}}_{U,I}^H{{\bf{w}}_I}\} \ge  \sqrt{{{\zeta}}}\sigma _{U},\tag{6b}\label{eq6b}\end{equation}
	\begin{equation} \kern 4pt \text{Im}\{{\bf{h}}_{U,I}^H{{\bf{w}}_I}\} =0,\kern 10pt\tag{6c}\label{eq6c}\end{equation}
	\begin{equation} \kern 21pt \text{Re}\{{\bf{h}}_{U,Q}^H{{\bf{w}}_Q}\} \ge  \sqrt{{{\zeta}}}\sigma _{U},\tag{6d}\label{eq6d}\end{equation}
	\begin{equation} \kern 4pt \text{Im}\{{\bf{h}}_{U,Q}^H{{\bf{w}}_Q}\} =0,\kern 6pt\tag{6e}\label{eq6e}\end{equation}
\end{subequations}
where ${{\zeta }}\in \mathbb{R}$ typically specifies the minimum SNR required for LU. Since the received SNR determines the probability of error, it is a main measure of communication quality. To recover correct symbols of PSK modulation at LU, the combination of two branches from transmitter I and Q should be in the same proportion. The constraints in \eqref{eq6b} and \eqref{eq6d} are to provide a prescribed quality communication assurance for LU, so that the received SNR of the in-phase and quadrature symbol components at LU are both greater than the specific SNR threshold ${{{\zeta}}}$. The constraints in \eqref{eq6c} and \eqref{eq6e} are to eliminate interference with the other branch. In addition, it is required that each transmitter should have a sufficiently large transmit power to guarantee that the link meets the received SNR requirements.

It can be easily found that the beamforming vectors of the distributed transmitters can be separately optimized. Then, the design of the beamforming vector of the transmitter I is given by
\begin{subequations}
\begin{equation} \kern10pt \textbf{P2:}  \kern 4pt \mathop {\min }\limits_{{{\bf{w}}_I}} \kern 2pt \Vert {\bf{w}}_I \Vert_2^2 \kern100pt \tag{7a} \label{eq7a}\end{equation}
\begin{equation}{\rm{s.t.}} \kern 2pt \text{Re}\{{\bf{h}}_{U,I}^H{{\bf{w}}_I}\} \ge  \sqrt{{{\zeta}}}\sigma _{U},\tag{7b}\label{eq7b}\end{equation}
\begin{equation} \kern 0pt \text{Im}\{{\bf{h}}_{U,I}^H{{\bf{w}}_I}\} =0.\kern 6pt\tag{7c}\label{eq7c}\end{equation}
\end{subequations}

In what follows, we will find the optimal solution of problem {\bf{P2}}. For the first step, we use ${\bf{h}}_{U,I}^H = {\rm{Re}}\{ {\bf{h}}_{U,I}^H\}  + j{\rm{Im}}\{ {\bf{h}}_{U,I}^H\}$ and ${{\bf{w}}_I} = {\rm{Re}}\{ {{\bf{w}}_I}\}  + j{\rm{Im}}\{ {{\bf{w}}_I}\}$ to separate the real and imaginary parts. Based on the following expression
\begin{eqnarray}
\begin{aligned}[b]
{\bf{h}}_{U,I}^H{{\bf{w}}_I} = &{\rm{Re}}\{ {\bf{h}}_{U,I}^H\} {\rm{Re}}\{ {{\bf{w}}_I}\}  - {\rm{Im}}\{ {\bf{h}}_{U,I}^H\} {\rm{Im}}\{ {{\bf{w}}_I}\}  \\&
+ j[{\rm{Im}}\{ {\bf{h}}_{U,I}^H\} {\rm{Re}}\{ {{\bf{w}}_I}\}  + {\rm{Re}}\{ {\bf{h}}_{U,I}^H\} {\rm{Im}}\{ {{\bf{w}}_I}\} ],
\end{aligned}
\label{eq8}\end{eqnarray}
then, we have
\begin{eqnarray}
	\begin{aligned}[b]
		\left\{ {\begin{array}{*{20}{l}}
				{{\rm{Re}}\{ {\bf{h}}_{U,I}^H{{\bf{w}}_I}\}  = {\bf{h}}_{I,1}^T{{{\bf{\tilde w}}}_I},}\\
				{{\rm{Im}}\{ {\bf{h}}_{U,I}^H{{\bf{w}}_I}\}  = {\bf{h}}_{I,2}^T{{{\bf{\tilde w}}}_I},}
		\end{array}} \right.
	\end{aligned}
	\label{eq9}\end{eqnarray}
where ${\bf{h}}_{I,1}^T = [{\rm{Re}}\{ {\bf{h}}_{U,I}^H\}$, $- {\rm{Im}}\{ {\bf{h}}_{U,I}^H\} ]$, ${\bf{h}}_{I,2}^T = [{\rm{Im}}\{ {\bf{h}}_{U,I}^H\}$, ${\rm{Re}}\{ {\bf{h}}_{U,I}^H\}]$, and ${{{\bf{\tilde w}}}_I} = {[{\rm{Re}}\{ {\bf{w}}_I^T\}, {\rm{Im}}\{ {\bf{w}}_I^T\} ]^T}$. Using \eqref{eq9} and the fact $\Vert {{{\bf{w}}_I}} \Vert_2^2 = \Vert {{{{\bf{\tilde w}}}_I}} \Vert_2^2$, problem {\bf{P2}} transforms into
\begin{subequations}
	\begin{equation}\textbf{P3:} \kern 4pt \mathop {\min }\limits_{{\bf{\tilde w}}_I} \kern 2pt \Vert {\bf{\tilde w}}_I \Vert_2^2 \tag{10a}\kern 66pt \label{eq10a}\end{equation}
	\begin{equation}{\rm{s.t.}} \kern 2pt  {\bf{h}}_{I,1}^T{{{\bf{\tilde w}}}_I} \ge  \sqrt{{{\zeta}}}\sigma _{U},\tag{10b}\label{eq10b}\end{equation}
	\begin{equation}\kern 0pt  {\bf{h}}_{I,2}^T{{{\bf{\tilde w}}}_I} =0.\kern 7pt  \tag{10c}\label{eq10c}\end{equation}
\end{subequations}
To eliminate the equality constraint \eqref{eq10c}, we use the singular value decomposition (SVD) operation on ${\bf{h}}_{I,2}^T$, which is expressed as
\begin{equation}
\begin{array}{l}
{\bf{h}}_{I,2}^T = {{{U}}_{I}}[\begin{array}{*{20}{c}}
{{{{\Sigma }}_{I}}}&{\boldsymbol{0}}
\end{array}]{[\begin{array}{*{20}{c}}
	{{\bf{V}}_{I}^{(1)}}&{{\bf{V}}_{I}^{(0)}}
	\end{array}]^T},
\label{eq11}\end{array}
\end{equation}
where ${{{U}}_{I}}\in \mathbb{R}^ {1 \times 1}$, ${{{{\Sigma }}_{I}}}$ is a singular value. According to the SVD property \cite{Matrix_Horn}, ${\bf{V}}_{I}^{(0)}\in \mathbb{R}^ {{2N_I} \times {(2N_I-1)}}$ is the last ${(2N_I-1)}$ right singular vectors corresponding to zero singular values, satisfying ${\bf{h}}_{I,2}^T{\bf{V}}_{I}^{(0)}={\textbf 0}$. We define ${\bf{B}}_I \buildrel \Delta \over = {\bf{V}}_{I}^{(0)}$. Since ${\bf{B}}_I{\bf{B}}_I^T={\bf{I}}_{2N_I}$, we have $\Vert {\bf{B}}_I{\boldsymbol{\xi}}_{I} \Vert_2^2=\Vert{\boldsymbol{\xi}}_{I} \Vert_2^2$ . By replacing ${\bf{\tilde w}}_{I}$ with ${\bf{B}}_I{\boldsymbol{\xi}}_{I}$, we consider the following equivalent problem with respect to ${\boldsymbol{\xi}}_{I}$
\begin{subequations}
	\begin{equation} \textbf{P4:} \kern 4pt \mathop {\min }\limits_{{\boldsymbol{\xi}}_{I}} \kern 2pt \Vert {\boldsymbol{\xi}}_{I} \Vert_2^2 \tag{12a}\kern 80pt \label{eq12a}\end{equation}
	\begin{equation}{\rm{s.t.}} \kern 2pt  {\bf{h}}_{I,1}^T{\bf{B}}_I{\boldsymbol{\xi}}_{I} \ge  \sqrt{{{\zeta}}}\sigma _{U}.\tag{12b}\label{eq12b}\end{equation}
\end{subequations}

Next, we employ an iterative algorithm to obtain the optimal solution of problem {\bf{P4}} with stationary convergence guarantee. To make such a problem tractable, we first add a real valued slack variable $\kappa  \ge 0$ to transform the inequality constraint in \eqref{eq12b} to an equality constraint as $ {\bf{h}}_{I,1}^T{\bf{B}}_I{\boldsymbol{\xi}}_{I} = \sqrt {\zeta}  {\sigma _U} + \kappa $. Using the penalty method \cite{Nonlinear_Bertsekas}, we then construct the following new problem
\begin{eqnarray}
\begin{aligned}[b]
\textbf{P5:} \kern 4pt \mathop {\min }\limits_{{\boldsymbol{\xi}}_{I},\kappa}\kern 2pt \Vert{\boldsymbol{\xi}}_{I} \Vert_2^2 + \lambda \vert{\bf{h}}_{I,1}^T{\bf{B}}_I{\boldsymbol{\xi}}_{I} - \sqrt {\zeta}  {\sigma _U}-\kappa\vert^2,
\end{aligned}
\label{eq13}\end{eqnarray}	
which is equivalent to problem {\bf{P4}} when $\lambda  \to +\infty $. To solve problem {\bf{P5}}, we resort an iterative algorithm by first tuning over $\kappa$ and keeping ${\boldsymbol{\xi}}_{I}$ fixed, and then optimizing ${\boldsymbol{\xi}}_{I}$ and viewing $\kappa$ fixed. Each iteration admits a closed-form solution. 

When we tune over $\kappa$ and consider ${\boldsymbol{\xi}}_{I}$ to be fixed, the optimization problem can be recast as
\begin{eqnarray}
\begin{aligned}[b]
\textbf{P6:} \kern 4pt \mathop {\min }\limits_{\kappa}\kern 2pt \vert{\bf{h}}_{I,1}^T{\bf{B}}_I{\boldsymbol{\xi}}_{I} - \sqrt {\zeta}  {\sigma _U}-\kappa\vert^2.
\end{aligned}
\label{eq14}\end{eqnarray}	
Considering $\kappa \ge 0$, the closed-form optimal solution of problem {\bf{P6}} is given by
\begin{eqnarray}
\begin{aligned}[b]
{{\kappa}}^\star = {\text{max}}\kern 2pt\{{\bf{h}}_{I,1}^T{\bf{B}}_I{\boldsymbol{\xi}}_{I} - \sqrt {\zeta}  {\sigma _U}, {{0}}\}.
\end{aligned}
\label{eq15}\end{eqnarray}
Next, we optimize ${{\boldsymbol{\xi}}_{I}}$ and view $\kappa$ fixed, and then the optimization problem is given by
\begin{eqnarray}
\begin{aligned}[b]
\textbf{P7:} \kern 4pt \mathop {\min }\limits_{{\boldsymbol{\xi}}_{I}}\kern 2pt \Vert {\boldsymbol{\xi}}_{I} \Vert_2^2 + \lambda  \vert{\bf{h}}_{I,1}^T{\bf{B}}_I{\boldsymbol{\xi}}_{I} - \sqrt {\zeta}  {\sigma _U}-\kappa\vert^2.
\end{aligned}
\label{eq16}\end{eqnarray}
The objective function of \eqref{eq16} can be constructed as
\begin{eqnarray}
\begin{aligned}[b]
f({\boldsymbol{\xi}}_{I}) =  \Vert {\boldsymbol{\xi}}_{I} \Vert_2^2 + \lambda \vert{\bf{h}}_{I,1}^T{\bf{B}}_I{\boldsymbol{\xi}}_{I} - \sqrt {\zeta}  {\sigma _U}-\kappa\vert^2.
\end{aligned}
\label{eq17}\end{eqnarray}
The optimal solution of objective function \eqref{eq17} can be obtained when the gradient of $f({\boldsymbol{\xi}}_{I})$ with respect to ${\boldsymbol{\xi}}_{I}$ equals to 0, i.e.,
\begin{align}
{\boldsymbol{\xi}}_{I}^\star =( \frac{{\bf{I}}_{2N-1}}{\lambda } +{\bf{B}}_I^T{\bf{h}}_{I,1}{\bf{h}}_{I,1}^T{\bf{B}}_I)^{ - 1}{\bf{B}}_I^T{\bf{h}}_{I,1}( \sqrt {\zeta}  {\sigma _U}+\kappa).
\label{eq18}\end{align}

In particular, the detailed procedures of the iterative algorithm to solve problem {\bf{P4}} is summarized in Algorithm 1.
\begin{table}[t]
	\begin{center}
		\begin{tabular}{llr}
			\hline
			\textbf{Algorithm 1: }Iterative algorithm for the problem {\bf{P4}}\\
			\hline
			\textbf{Initialization:} Pick up feasible points ${{\boldsymbol{\xi}}_{I}^0} \in \mathbb{R}^{(2N-1) \times 1}$,\\ 
			\kern 10pt $\lambda \in [0, + \infty )$, $r=0$, and tolerance $\epsilon>0$; \\
			1. \textbf{Loop}\\
			2. \kern 10pt $r=r+1$;\\
			3. \kern 10pt 	Determine ${\kappa^{r-1}}$ by substituting ${\boldsymbol{\xi}}_{I}^{r-1}$ into \eqref{eq15};\\
			4. \kern 10pt   Determine ${\boldsymbol{\xi}}_{I}^{r}$ by substituting $\kappa^{r-1}$ into \eqref{eq18}; \\
			5. \kern 10pt   Stop if $\Vert {\boldsymbol{\xi}}_{I}^{r}-{\boldsymbol{\xi}}_{I}^{r-1} \Vert_2^2<\epsilon$; \\
			6. \textbf{End loop}  \\
			\textbf{Output:} Get the finally optimal solution  ${\boldsymbol{\xi}}_{I}$; \\
			\hline
		\end{tabular}
		\label{tab1}
	\end{center}
\end{table}

In what follows, we will prove that the optimization problem is solved with stationary convergence by the iterative algorithm. Assume that ${{\boldsymbol{\xi}}_{I}^0}$ and ${{\kappa}^0}$ are initial values. According to \eqref{eq15} and \eqref{eq18}, we can obtain the optimal solutions ${{\boldsymbol{\xi}}_{I}^\star}$ and ${\kappa}^\star$, which satisfy
\begin{eqnarray}
\begin{aligned}[b]
f({\boldsymbol{\xi}}_{I}^\star,{{\kappa}^{\star}}) \le f({\boldsymbol{\xi}}_{I}^0,{\kappa}^\star) \le f({\boldsymbol{\xi}}_{I}^0,{\kappa}^{0}).
\end{aligned}
\label{eq19}\end{eqnarray}
It can be noted that fixing ${{\boldsymbol{\xi}}_{I}}$ in {\bf{P6}}, or $\kappa$ in {\bf{P7}}, lead into a convex function, and each iteration in Algorithm 1 monotonically approaches to the optimal value since the object function $f({\boldsymbol{\xi}}_{I},{{\kappa}})$ is lower bounded at zero. Therefore, we establish the stationary convergence guarantee for the iterative algorithm to find the optimal solution. After obtaining the optimal solution ${\boldsymbol{\xi}}_{I}^\star$, the optimal beamforming vector ${\bf{w}}_I^\star$ can be found.

With similar manipulations, the optimization problem for design of the beamforming vector at transmitter Q can be formulated as
\begin{subequations}
	\begin{equation}\kern 20pt \textbf{P8:} \kern 4pt  \mathop {\min }\limits_{{{\bf{w}}_Q}} \kern 2pt \Vert {\bf{w}}_Q \Vert_2^2 \kern 90pt\tag{20a} \label{eq20a}\end{equation}
	\begin{equation}{\rm{s.t.}} \kern 2pt \text{Re}\{{\bf{h}}_{U,Q}^H{{\bf{w}}_Q}\} \ge  \sqrt{{{\zeta}}}\sigma _{U},\tag{20b}\label{eq20b}\end{equation}
	\begin{equation} \kern 0pt \text{Im}\{{\bf{h}}_{U,Q}^H{{\bf{w}}_Q}\} =0.\kern 6pt\tag{20c}\label{eq20c}\end{equation}
\end{subequations}
Using a similar approach as {\bf{P2}}, we can obtain the optimal beamforming vector ${\bf{w}}_Q^\star$.

\subsection{AN Projection Matrix Design}
It may be possible for a sufficiently sensitive Eve to intercept confidential information from the message power that leaks through proximal mainlobe or sidelobes. Consequently, it is of interest to add AN to interfere with Eves. As the location information of Eves is unavailable at the transmitter, Eves potentially exist in anywhere. When the location of Eve satisfies $2\pi {f_c}(\frac{{{r_{E,Q}}}-{{r_{E,I}}}}{c}+ \tau)+{\varphi _{E,I}}-\varphi _{E,Q}=2k\pi, k=0, \pm 1,\pm 2,... $ with ${\varphi _{E,I}}={\text {arg}}\{{\bf{h}}_{E,I}^H{{\bf{w}}_I}\}$ and $\varphi _{E,Q}={\text {arg}}\{{\bf{h}}_{E,Q}^H{{\bf{w}}_Q}\}$, the carriers of the received branches are orthogonal, which causes no inter-branch interference. At this time, AN-aided beamforming poses interference outside the mainbeam to prevents eavesdropping.

We design the AN at the transmitter I whose elements are equal to ${{\mathbf{n}}_{A,I}}={{\bf{P}}_I}{{\bf{z}}_I}$, where ${{\bf{P}}_I}\in \mathbb{C}^{{N_I} \times {N_I}}$ denotes the AN projection matrix for imposing interference to Eves, and AN vector ${\mathbf{z}}\in \mathbb{C}^{{N_I} \times 1}$ consists of complex Gaussian variables with zero-mean and unit-variance, satisfying ${\bf{z}}\sim{\mathcal{CN}}({{{0}}},{{\bf{I}}_{N_I}})$. We view all directions outside the LU's mainlobe as the possible directions of Eves. Accordingly, we define the set of undesired directions as
\begin{eqnarray}
	\begin{aligned}[b]
		{{\mathcal{D}} _{E,I}}= [ - \frac{\pi }{2},\frac{\pi }{2}]\backslash [{\theta _{U,I}} - \frac{{\theta _I^{BW}}}{2},{\theta _{U,I}} + \frac{{\theta _I^{BW}}}{2}],
	\end{aligned}
	\label{eq21}\end{eqnarray}
where $\theta _I^{BW} = \frac{{50.7^\circ c}}{{{f_c}{N_I}{d_I}}}$ is the half-power main-lobe beamwidth of transmitter I \cite{Antennas_Kraus}.

As is well known, the add of AN is able to dynamically interfere with Eves, but it also consumes part of total transmit power. By taking the side-lobe message power into account, the proposed AN design makes full use of the limited transmit AN power to achieve the target SINR value in the undesired directions. With regard to this, the goal of AN projection matrix design is to minimize AN power consumption, while satisfying the target SINR constraint for all undesired directions, as well as the null space of the LU to cancel the interference caused by the AN to the LU, which is given by$\footnote{{If the noise variance of Eves cannot be obtained due to the unavailable information of passive Eves, the constraint can be converted to signal-to-interference ratio (SIR).}}$
\begin{subequations}
	\begin{equation}  \kern 1pt \textbf{P9:} \kern 4pt  \mathop {\min }\limits_{{\bf{P}}_I}\kern 2pt \mathbb{E}\{\vert {{\bf{P}}_I}{\bf{z}}_I\vert^2 \}\kern 160pt \tag{22a}\label{eq22a}\end{equation}
	\begin{equation} \kern 0pt {\rm{s.t.}}\kern 2pt\mathbb{E}\{\vert {\bf{h}}_{U,I}^T{{\bf{P}}_I}{\bf{z}}_I \vert_2^2\} = 0,\kern 76pt \tag{22b}\label{eq22b}\end{equation}
	\begin{equation}\kern 25pt \frac{{\vert{\bf{h}}_{E,I}^T{{\bf{w}}_I}\vert^2}}{{ \mathbb{E}\{\vert{\bf{h}}_{E,I}^T{{\bf{P}}_I}{\bf{z}}_I\vert_2^2\} + \sigma _E^2}} \le \gamma, \forall {\theta _{E,I}} \in {{\mathcal{D}} _{E,I}},\tag{22c}\label{eq22c}\end{equation}
\end{subequations}	
where $\gamma$ is the predefined maximum SINR toleration for Eves, and ${{\mathcal{D}}_{E,I}}$ denotes the set of undesired directions in which Eves may potentially exist. The constraint in \eqref{eq22b} implies that AN is interference-free with LU, and constraint in \eqref{eq22c} represents that received SINRs at potential Eves are less than target SINR value. The design of AN not only achieves the target SINR for the undesired directions but also reduces AN power consumption. Based on the fact $\mathbb{E}\{{\bf{z}}_I{\bf{z}}_I^H\}={\bf{I}}_{N_I}$, problem {\bf{P9}} can be converted as
\begin{subequations}
	\begin{equation} \kern 0pt\textbf{P10:} \kern 4pt \mathop {\min }\limits_{{\bf{P}}_I}\kern 2pt {\rm{  tr}}({{\bf{P}}_I}{{\bf{P}}_I^T})\kern 106pt \tag{23a}\label{eq23a}\end{equation}
	\begin{equation} \kern 25pt {\rm{s.t.}} \kern 2pt \Vert {\bf{h}}_{U,I}^T{{\bf{P}}_I} \Vert_2^2 = 0, \kern 80pt \tag{23b}\label{eq23b}\end{equation}
	\begin{equation} \kern 45pt \frac{\vert{\bf{h}}_{E,I}^T{{\bf{w}}_I}\vert^2}{{\Vert{\bf{h}}_{E,I}^T{{\bf{P}}_I}\Vert_2^2 + \sigma _E^2}} \le \gamma, \forall {\theta _{E,I}} \in {{\mathcal{D}} _{E,I}}. \tag{23c}\label{eq23c}\end{equation}
\end{subequations}	
We use the fact that ${\bf{h}}_{U,I}^T{{\bf{P}}_I}{{\bf{P}}_I^T}{{\bf{h}}_{U,I}} ={\rm{tr}}({\bf{h}}_{U,I}^T{{\bf{P}}_I}{{\bf{P}}_I^T}$ ${{\bf{h}}_{U,I}}) ={\rm{tr}}({{\bf{h}}_{U,I}}{\bf{h}}_{U,I}^T{{\bf{P}}_I}{{\bf{P}}_I^T})$, and ${\bf{h}}_{E,I}^T{{\bf{P}}_I}{{\bf{P}}_I^T}$ ${{\bf{h}}_{E,I}} = {\rm{tr}}({\bf{h}}_{E,I}^T{{\bf{P}}_I}{{\bf{P}}_I^T}{{\bf{h}}_{E,I}}) = {\rm{tr}}({{\bf{h}}_{E,I}}{\bf{h}}_{E,I}^T{{\bf{P}}_I}{{\bf{P}}_I^T})$, and then define ${{\bf{\Gamma }}_I} \buildrel \Delta \over = {\bf{P}}_I{{\bf{P}}_I^T}$, ${{\bf{\Pi }}_{U,I}} \buildrel \Delta \over = {{\bf{h}}_{U,I}}{\bf{h}}_{U,I}^T$, and ${{\bf{\Pi }}_{E,I}} \buildrel \Delta \over = {{\bf{h}}_{E,I}}{\bf{h}}_{E,I}^T$. Now, problem {\bf{P10}} with respect to the AN projection matrix can be converted as the following equivalent problem with respect to AN covariance matrix
\begin{subequations}
	\begin{equation} \kern 24pt \textbf{P11:} \kern 4pt \mathop {\min }\limits_{{\bf{\Gamma }}_I}\kern 2pt {\rm{  tr}}({{\bf{\Gamma }}_I})\kern 105pt \tag{24a}\label{eq24a}\end{equation}
	\begin{equation}{\rm{s.t.}} \kern 2pt{\rm{tr}}({{\bf{\Pi }}_{U,I}}{{\bf{\Gamma }}_I}) = 0,\kern 37pt\tag{24b}\label{eq24b}\end{equation}
	\begin{equation} \kern 62pt {\rm{ tr}}({{\bf{\Pi}}_{E,I}}{{\bf{\Gamma }}_I}) \ge {\tilde \gamma }, \forall {\theta _{E,I}} \in {{\mathcal{D}} _{E,I}}, \tag{24c}\label{eq24c}\end{equation}
	\begin{equation} \kern 4pt {{\bf{\Gamma }}_I} \succeq 0,\kern 62pt  \tag{24d}\label{eq24d}\end{equation}
\end{subequations}		
where ${\tilde \gamma } = |{\bf{h}}_{E,I}^T{{\bf{w}}_I}{|^2}/\gamma - \sigma _E^2$, and the inequality in \eqref{eq24d} implies that AN covariance matrix is symmetric positive semidefinite \cite{Transmit_Sidiropoulos}. Then, we vectorize the problem {\bf{P11}}, which can be expressed as
\begin{subequations}
	\begin{equation} \kern 0pt\textbf{P12:} \kern 4pt \mathop {\min }\limits_{{\bf{\Gamma }}_I} \kern 2pt {\rm{vec}}({{\bf{I}}_{{N_I}}})^T{\rm{vec}}({{\bf{\Gamma }}_I}) \kern 78pt\tag{25a}\label{eq25a}\end{equation}
	\begin{equation} \kern 16pt{\rm{s.t.}} \kern 2pt{\rm{vec}}({\bf{\Pi }}_{U,I}^T)^T{\rm{vec}}({{\bf{\Gamma }}_I}) = 0,\kern 40pt \tag{25b}\label{eq25b}\end{equation}
	\begin{equation} \kern 52pt{\rm{vec}}({\bf{\Pi }}_{E,I}^T)^T{\rm{vec}}({{\bf{\Gamma }}_I}) \ge {\tilde \gamma }, \forall {\theta _{E,I}} \in {{\mathcal{D}} _{E,I}},\kern 0pt \tag{25c}\label{eq25c}\end{equation}
	\begin{equation} \kern 0pt {{\bf{\Gamma }}_I} \succ 0,\kern 104pt \tag{25d}\label{eq25d}\end{equation}
\end{subequations}
The important observation is that the optimization problem {\bf{P12}} is in a form suitable for semidefinite relaxation (SDR) \cite{Quasi_Wing}, which can be solved by utilizing interior point methods. Optimization tool, such as CVX \cite{CVX_Grant}, is particularly efficient for such a SDR problem. After obtaining the optimal AN covariance matrix of problem {\bf{P12}}, the AN can be calculated via eigendecomposition of ${\bf{\Gamma }}_I^\star={{\bf{U}}_{{{\Gamma }}_I} }{{\boldsymbol{\Sigma}} _{{{\Gamma }}_I} }{\bf{U}}_{{{\Gamma }}_I} ^T$, and choose AN vector ${\bf{z}}_I$ such that ${{\bf{n}}_{A,I}} = {{\bf{U}}_{{{\Gamma }}_I}  }{\bf{\Sigma }}_{{{\Gamma }}_I}  ^{1/2}{{\bf{z}}_I}$. 

Next, we can calculate the AN of transmitter Q in a similar manner.

\textsl{Remark 3:} We should mention that our design is to seek to minimize the total transmit power (including message power and AN power) by optimizing the beamforming vector and the AN projection matrix subject to meeting certain constraints. On other hand, our proposed scheme is also available for the case of the total transmit power constraint. More specifically, after obtaining the optimal beamforming vectors and AN projection matrix by the proposed scheme, the extra transmit power is then assigned to AN to enhance interference with Eves.
\subsection{Synchronization of Signals}

It is important to note that the in-phase and quadrature branches from two transmitters should arrive at LU synchronously. Considering the different range between LU and two transmitters, we adjust the delay at transmitter Q to guarantee the synchronization. On one hand, to guarantee that the symbol components arrived at LU are aligned, transmitter Q needs to adjust the latency time to compensate for the path delay difference. When ${r_{U,I}} > {r_{U,Q}}$ (or ${r_{U,I}} < {r_{U,Q}}$), transmitter Q delays (or advances) the symbol emission by 
\begin{eqnarray}
	\begin{aligned}[b]
		\tau  = ({r_{U,I}} - {r_{U,Q}})/c.
	\end{aligned}
	\label{eq26}\end{eqnarray}
On the other hand, to guarantee mutually orthogonal carriers, the carrier phase of transmitter Q needs to be adjusted to compensate for the phase delay difference as
\begin{eqnarray}
	\begin{aligned}[b]
		\delta  = {e^{-j2\pi {f_c}\tau}}.
	\end{aligned}
	\label{eq27}\end{eqnarray} 

\subsection{Extension of Multiuser System}
So far, we have provided secure transmission for the case of single LU. In practice, it usually requires that multiple LUs simultaneously receive their own individual confidential message streams. We now extend the proposed AN-aided D3M scheme to the case of multiuser. Since multiple message streams simultaneous transmission needs to be supported, an improved scheme is required. To do so, we employ multi-beam DM and combat the intersymbol interference among multiple users. Multi-beam DM is capable of transmitting multiple message streams to corresponding LUs in different directions simultaneously while distorting the constellations in all other directions \cite{Optimal_Chukhno}. 

Without loss of generality, $K$ LUs are considered in this system, which are located at $({r_{U_k,I}},{\theta_{U_k,I}})$ relative to transmitter I, and $({r_{U_k,Q}},{\theta_{U_k,Q}})$ relative to transmitter Q, for the LU $k$, $\forall k \in \mathcal{K}$, $\mathcal{K} \buildrel \Delta \over  = \left\{ {1,2,...,K} \right\}$. Let us define the channel matrix of the transmitter I and Q for all LUs as
\begin{eqnarray}
	\begin{aligned}[b]
		\left\{ {\begin{array}{*{20}{l}}
				{{{\bf{H}}_{U,I}} \buildrel \Delta \over = [{{\bf{h}}_{{U_1},I}},{{\bf{h}}_{{U_2},I}},...,{{\bf{h}}_{{U_K},I}}],}\\
				{{{\bf{H}}_{U,Q}}\buildrel \Delta \over  = [{{\bf{h}}_{{U_1},Q}},{{\bf{h}}_{{U_2},Q}},...,{{\bf{h}}_{{U_K},Q}}],}
		\end{array}} \right.
	\end{aligned}
	\label{eq28}\end{eqnarray}
where ${{\bf{h}}_{{U_k},I}} \buildrel \Delta \over = {{\bf{h}}_I}({r_{{U_k},I}},{\theta _{{U_k},I}})$ and ${{\bf{h}}_{{U_k},Q}} \buildrel \Delta \over = {{\bf{h}}_Q}({r_{{U_k},Q}},{\theta _{{U_k},Q}})$ are the channel vector from the transmitter I and Q to the LU $k$, $\forall k \in \mathcal{K}$, respectively. Then, the multi-beam transmit signals at transmitter I and Q can be given by
\begin{eqnarray}
\begin{aligned}[b]
\left\{ {\begin{array}{*{20}{l}}
		{{{\bf{s}}_I}(t) =\sum\limits_{k = 1}^K {\bf{w}}_{k,I}{x_{k,I}}(t) + {{\bf{n}}_{A,I}},}\\
		{{{\bf{s}}_Q}(t) = \sum\limits_{k = 1}^K \delta_k \cdot {\bf{w}}_{k,Q}{x_{k,Q}}(t-\tau_k) + {{\bf{n}}_{A,Q}},}
\end{array}} \right.
\end{aligned}
\label{eq29}\end{eqnarray}
where ${x_{k,I}}(t)\in \mathbb{R}$ and ${x_{k,Q}}(t)\in \mathbb{R}$ are the in-phase and quadrature symbol components toward LU $k$, after deposition of the input message ${x}_k(t)\in \mathbb{C}$ of the LU $k$, $\forall k \in \mathcal{K}$, ${x}_k(t)={x_{k,I}}(t)+j {x_{k,Q}}(t)$, with $\mathbb{E}\{{\left| {x_{k}}(t) \right|^2}\} = 1$, ${\bf{w}}_{k,I}\in \mathbb{C}^{N_I \times 1}$ is the beamforming vector at transmitter I for processing the in-phase signal component ${x_{k,I}}(t)$ to the LU $k$, and ${\bf{w}}_{k,Q}\in \mathbb{C}^{N_I \times 1}$ is the beamforming vector at transmitter Q for processing quadrature signal component ${x_{k,Q}}(t)$ to the LU $k$,  $\forall k \in \mathcal{K}$.

After passing through wireless channels, the received signal vector of LUs can be written as
\begin{align}
{{\bf{y}}_U}(t)\! = \! {\rm{Re}}\{ {{\bf{G}}_{U,I}}{\bf{H}}_{U,I}^H{{\bf{s}}_I}(t)\}\! +\! {\rm{Im}}\{ {{\bf{G}}_{U,Q}}{\bf{H}}_{U,Q}^H {{\bf{s}}_Q}(t)\}\! +\! {{\bf{n}}_U},
\label{eq30}\end{align}	
where ${{\bf{G}}_{U,I}} = {\rm{diag}}{({\bf{g}}_{U,I})}$, ${{\bf{g}}_{U,I}} = [{{\rm{g}}_{{U_1},I}},{{\rm{g}}_{{U_2},I}},...,{{\rm{g}}_{{U_K},I}}]^T$ is the carrier vector of transmitter I with ${g_{U_k,I}} = {e^{j2\pi {f_c}(t - \frac{{{r_{U_k,I}}}}{c})}}$, $\forall k \in \mathcal{K}$, ${{\bf{G}}_{U,Q}} ={\rm{ diag}}({\bf{g}}_{U,Q})$, ${{\bf{g}}_{U,Q}}= [{{\rm{g}}_{{U_1},Q}},{{\rm{g}}_{{U_2},Q}},...,{{\rm{g}}_{{U_K},Q}}]^T$ is the carrier vector of transmitter Q with ${g_{U_k,Q}} = {e^{j2\pi {f_c}(t - \frac{{{r_{U_k,Q}}}}{c})}}$, $\forall k \in \mathcal{K}$, and ${{\bf{n}}_{U}(t)}=[{{n_{U_1}}(t)},{{n_{U_2}}(t)},...,{{n_{U_K}}(t)}]^T$ is the AWGN vector at LUs with ${{{{n}}_{U_k}}}\sim{\mathcal{N}}({0},\sigma _{U_k}^2)$, $\forall k \in \mathcal{K}$. Thereinto, the received signal at the LU $k$, $\forall k \in \mathcal{K}$, is given in \eqref{eq31}, shown at the bottom of the page,
where $\tau_k  = ({r_{U_k,I}} - {r_{U_k,Q}})/c$ is the latency factor , and $\delta_k  = {e^{-j2\pi {f_c}\tau_k}}$ is the phase compensation factor, for the LU $k$, $\forall k \in \mathcal{K}$.
\begin{figure*}[b]
	\hrulefill %
	\begin{align}\label{eq31}
		& {y_{{U_k}}}(t) = {\rm{Re}}\{ {g_{{U_k},I}}{\bf{h}}_{{U_k},I}^H{{\bf{w}}_{k,I}}{x_{k,I}}(t)\}  + {\rm{Re}}\{ {g_{{U_k},I}}{\bf{h}}_{{U_k},I}^H\sum\limits_{i = 1,i \ne k}^K {{{\bf{w}}_{i,I}}{x_{i,I}}(t)} \}  + {\rm{Re}}\{ {g_{{U_k},I}}{\bf{h}}_{{U_k},I}^H{{\bf{n}}_{A,I}}\}\notag 
		\\&  \kern 42pt
		+ {\rm{Im}}\{ {g_{{U_k},Q}}{\delta _k}{\bf{h}}_{{U_k},Q}^H{{\bf{w}}_{k,Q}}{x_{k,Q}}(t - {\tau _k})\}  + {\rm{Im}}\{ {g_{{U_k},Q}}{\bf{h}}_{{U_k},Q}^H\sum\limits_{i = 1,i \ne k}^K {{\delta _i}{{\bf{w}}_{i,Q}}{x_{i,Q}}} (t - {\tau _i})\}	\notag
		\\& \kern 42pt
		+ {\rm{Im}}\{ {g_{{U_k},Q}}{\bf{h}}_{{U_k},Q}^H{{\bf{n}}_{A,Q}}\}  + {n_{{U_k}}},\tag{31}
	\end{align}
\end{figure*}

For multi-LU secure transmission, we aim to simultaneously transmit multiple interference-free symbol streams to multiple LUs. For notational clarity, we define complement of channel matrix of the LU $k$, $\forall k \in \mathcal{K}$, as
\begin{eqnarray}
\setcounter{equation}{32}
\begin{aligned}[b]
{{\bf{H}}_{ - {k},I}} \buildrel \Delta \over = [{{\bf{h}}_{{U_1},I}},...,{{\bf{h}}_{{U_{k - 1}},I}},{{\bf{h}}_{{U_{k + 1}},I}}...,{{\bf{h}}_{{U_K},I}}].
\end{aligned}
\label{eq32}\end{eqnarray} 
Then, for the case of multiuser, the design of beamforming vector $k$, $\forall k \in \mathcal{K}$, at transmitter I, is given by
\begin{subequations}
\begin{equation} \textbf{P12:} \kern 4pt \mathop  {\min }\limits_{{\bf{w}}_{k,I}} \kern 2pt \Vert {\bf{w}}_{k,I} \Vert_2^2 \kern 82pt \tag{33a}\label{eq33a}\end{equation}
\begin{equation}{\rm{s.t.}} \kern 2pt \text{Re}\{{\bf{h}}_{{U_k},I}^H{\bf{w}}_{k,I}\} \ge  \sqrt{{{\zeta}_k}}\sigma _{U_k},\tag{33b}\label{eq33b}\end{equation}
\begin{equation}\kern 0pt \text{Im}\{{\bf{h}}_{{U_k},I}^H{\bf{w}}_{k,I}\} =0,\kern 14pt\tag{33c}\label{eq33c}\end{equation}
\begin{equation}\kern 0pt {\bf{H}}_{-{k},I}^H{\bf{w}}_{k,I} = {\textbf{0}},\kern 32pt\tag{33d}\label{eq33d}\end{equation}
\end{subequations}
where ${{\zeta}_k}$ is the required SNR of received signal at the LU $k$, $\forall k \in \mathcal{K}$. Similarly, the constraints in \eqref{eq33b} and \eqref{eq33c} are to preserve the validity transmission of in-phase signal component, so that the received SNR at the LU $k$ is more than a given SNR threshold ${{\zeta}_k}$, $\forall k \in \mathcal{K}$. The constraint in \eqref{eq33d} imposes the strict constraint on the inter-users message interference by forcing the message for the LU $k$ to transmit through the null spaces of all remaining LUs. Similar to the processes of single LU, we separate the real and imaginary parts as  ${{{\bf{\tilde w}}}_{k,I}} = [{\rm{Re}}\{ {\bf{w}}_{k,I}^T\}$, ${\rm{Im}}\{ {\bf{w}}_{k,I}^T\} ]^T$, ${\bf{h}}_{k,I,1}^T =[{\rm{Re}}\{ {\bf{h}}_{U_k,I}^H\}$, $- {\rm{Im}}\{ {\bf{h}}_{U_k,I}^H\} ]$, ${\bf{h}}_{k,I,2}^T = [{\rm{Im}}\{ {\bf{h}}_{U_k,I}^H\}$, ${\rm{Re}}\{ {\bf{h}}_{U_k,I}^H\} ]$, ${\bf{H}}_{{-k,I,1}}^T = [{\rm{Re}}\{ {\bf{H}}_{-{k},I}^H\}$, $ - {\rm{Im}}\{{\bf{H}}_{-{k},I}^H\} ]$, and  ${\bf{ H}}_{{-k,I,2}}^T = [{\rm{Im}}\{ {\bf{H}}_{-{k},I}^H\}$, ${\rm{Re}}\{{\bf{H}}_{-{k},I}\}^H]$.
By stacking the constraints, we can encapsulate the problem {\bf{P12}} as
\begin{subequations}
\begin{equation} \kern 17pt \textbf{P13:} \kern 4pt \mathop {\min }\limits_{{\bf{w}}_{k,I}} \kern 2pt \Vert {\bf{\tilde w}}_{k,I} \Vert_2^2 \kern 80pt \tag{34a} \label{eq34a}\end{equation}
\begin{equation}{\rm{s.t.}} \kern 2pt {\bf{h}}_{{k,I,1}}^T{\bf{\tilde w}}_{I,k} \ge  \sqrt{{{\zeta}_k}}\sigma _{U_k},\tag{34b}\label{eq34b}\end{equation}
\begin{equation}\kern 0pt {\bf{\tilde H}}_{{k,I}}^T{\bf{\tilde w}}_{I,k} =\textbf{0},\kern 18pt\tag{34c}\label{eq34c}\end{equation}
\end{subequations}
where ${{{\bf{\tilde H}}}_{k,I}} = [{{\bf{H}}_{ - k,I,1}},{{\bf{H}}_{ - k,I,2}},{{\bf{h}}_{k,I,2}}]$. In consequence, the form of problem {\bf{P13}} is the same as that of problem {\bf{P2}}. To make problem feasible, it is requested to more number of transmit antennas than the number of LUs, i.e., $N_I>K$. By using the SVD operation on ${\bf{\tilde H}}_{{k,I}}^T\in \mathbb{R}^ {{(2K-1)} \times {2N_I}}$ to eliminate the equality constraint. Then, we employ an iterative algorithm to obtain the optimal solution of problem {\bf{P13}} similar as {\bf{P3}}.

For the design of AN, in the case of multiuser, it is the same as the case of single LU, except for a slight difference in constraint on LUs; that is, the constraint is replaced with  ${{\bf{\Pi }}_{U,I}} \buildrel \Delta \over = {{\bf{H}}_{U,I}}{\bf{H}}_{U,I}^T$ to eliminate the interference with all LUs. 

\section{Analysis of Performance}
\subsection{Secrecy Performance}
The received radio frequency (RF) signal emitted by two distributed array transmitters with the designed beamforming vectors and AN at LU, is given by
\begin{align}
	{y_U}(t) =& \sqrt{{{\zeta}}}\sigma _{U}\cos[{2\pi {f_c}(t - \frac{{{r_{U,I}}}}{c})}]{x_I}(t) \notag\\ 
	&	+ \sqrt{{{\zeta}}}\sigma _{U}\sin[{2\pi {f_c}(t - \frac{{{r_{U,I}}}}{c})}]{x_Q}(t).
	\label{eq35}\end{align}
A normal phase-demodulation receiver, under perfect phase and carrier recovery, obtains the in-phase and quadrature symbol components baseband signals by down-convert operation and low-pass filter as follows \cite[Ch. 6]{Wireless_Goldsmith}:
\begin{align}
	{\left\{{\begin{array}{*{20}{l}}
				{y_U}(t) \cdot \cos [2\pi {f_c}(t\! -\! \frac{{{r_I}}}{c})] 	\overset{\rm{LP}}\to {y_{U,I}}(t)\! =\! \sqrt {\zeta } {\sigma _U} \cdot {x_I}(t)\! +\! {n_U},\\
				{y_U}(t) \cdot \sin [2\pi {f_c}(t\! -\! \frac{{{r_I}}}{c})] 	\overset{\rm{LP}}\to {y_{U,Q}}(t)\! =\! \sqrt {\zeta } {\sigma _U} \cdot {x_Q}(t)\! +\! {n_U}.
		\end{array}} \right.}
	\label{eq36}\end{align}	
Since $|x(t)|^2=1$, we can see that the received power is $ \zeta  {\sigma _U^2} |x_I(t)|^2+ \zeta  {\sigma _U^2} |x_Q(t)|^2$, which means that the received signal at LU meets the received SNR requirement. The original message $x(t)$ can be obtained by the combination of the in-phase and quadrature symbol components. The transmitters are well-designed for the secure transmission such that LU can directly recover messages. One can see that no extra signal processing is imposed at LU. Accordingly, the proposed scheme provides an extremely low-complexity structure for LU. That is, just a standard PSK receiver can effectively work, and hence the proposed scheme can effectively facilitate its application.

From the perspective of wiretapping, our design imposes  much harder for Eves to wiretap the confidential messages compared to conventional DM techniques realized by a single transmitter. Next, we present the benefits in terms of the security enhancement. 

a) Decomposed and distributed transmission structure. Since directions of Eve relative to transmitter I and Q cannot both be the same as that of the LU, , i.e., ${\theta _{E,I}} \ne {\theta _{U,I}}$, or ${\theta _{E,Q}} \ne {\theta _{U,Q}}$, as per \eqref{eq5}, the item ${\bf{h}}_{E,I}^H{{\bf{w}}_I} \notin \mathbb{R}$ or ${\bf{h}}_{E,Q}^H{{\bf{w}}_Q} \notin \mathbb{R}$, then in-phase branch will interfere with quadrature branch and vice versa. 

b) Directivity of array beamforming. For the proposed scheme, the transmitted signal is decomposed into two mutually orthogonal symbol branches transmitted by distributed transmitters. Only the received signal within the mainlobe has a high SNR. The effective receiving zone only occurs in the intersection of the two  main-lobe beamforming.

c) Spatial path delay. Due to the different range from transmitter I and Q to Eves, the transmission path delay is different. This steers the carriers of two branches to lose orthogonality, and thus incurs incorrect recovery happening at Eves. What's more, if the path delay difference is over a symbol period, it may even cause symbol misalignment. 

\begin{figure}
	\centering
	\includegraphics[width=0.8\columnwidth]{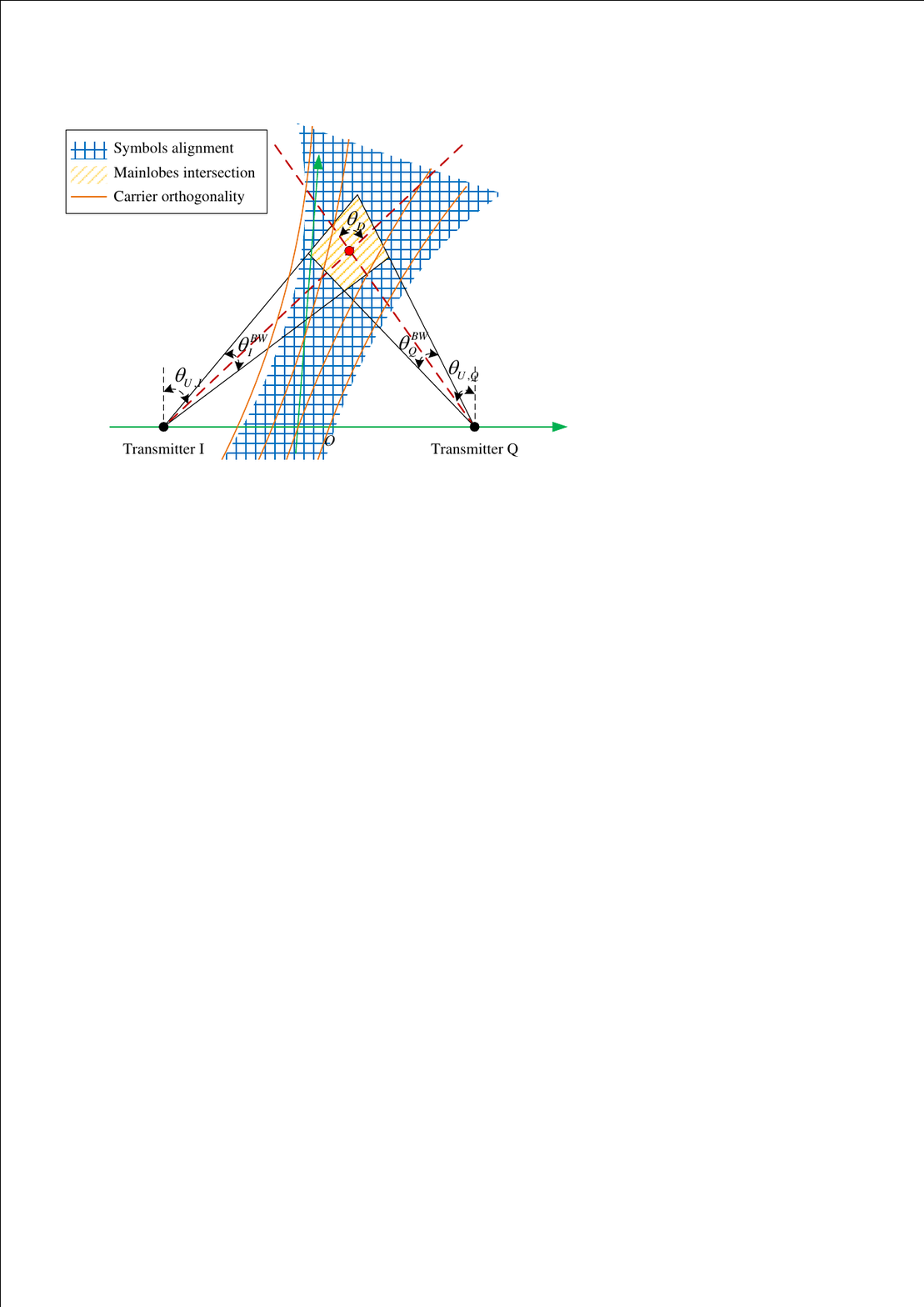}
	\caption{Geometric illustration of the effective receiving zone.}
	\label{fig3}
\end{figure}

We utilize the decomposed and distributed transmission structure, directivity of array beamforming, and the spatial path delay to make it more difficult for Eves to intercept information, and thus enhance the security. The intersection of mainlobes, symbols alignment, and carrier orthogonality, all together form the effective receiving zone, as shown in Fig \ref{fig3}.

In the previous section, we establish the strategy for PLS of wireless communications by optimizing the beamforming vectors and AN, which is capable of transmitting confidential messages to LU with a weak message leakage and strong interference outside the neighborhood around the desired location. Below, we will analyze the effectiveness of the proposed scheme in terms of average SINR, average SER, and average secrecy rate, which are considered as secrecy metrics.

The messages can be recovered as long as both of the in-phase and quadrature symbol components are correct. Therefore, we choose the maximum SINR for being conducive to the recovery of the symbol components. For notational convenience, we drop the arguments of $({r_I},{\theta _I})$ and $({r_Q},{\theta _Q})$ in ${\bf{h}}_I$ and ${\bf{h}}_Q$ when there is no ambiguity. For receivers located at arbitrary locations, the in-phase symbol components may be carried at the in-phase branch or quadrature branch, i.e.,
\begin{align}
	{\left\{{\begin{array}{*{20}{l}}
			{\Gamma _1^{{x_I}} = \frac{{{\rm{R}}{{\rm{e}}^2}\{ {\bf{h}}_I^H{{\bf{w}}_I}\} }}{{{\rm{I}}{{\rm{m}}^2}\{ {\delta _\Delta }{\bf{h}}_Q^H{{\bf{w}}_Q}\}  + {\rm{R}}{{\rm{e}}^2}\{ {\bf{h}}_I^H{{\bf{n}}_{A,I}}\}  + {\rm{I}}{{\rm{m}}^2}\{ {\delta _E}{\bf{h}}_Q^H{{\bf{n}}_{A,Q}}\}  + {\sigma ^2}}},}\\
			{\Gamma _2^{{x_I}} = \frac{{{\rm{I}}{{\rm{m}}^2}\{ {\bf{h}}_I^H{{\bf{w}}_I}\} }}{{{\rm{R}}{{\rm{e}}^2}\{ {\delta _\Delta }{\bf{h}}_Q^H{{\bf{w}}_Q}\}  + {\rm{I}}{{\rm{m}}^2}\{ {\bf{h}}_I^H{{\bf{n}}_{A,I}}\}  + {\rm{R}}{{\rm{e}}^2}\{ {\delta _E}{\bf{h}}_Q^H{{\bf{n}}_{A,Q}}\}  + {\sigma ^2}}},}
		\end{array}} \right.}
	\label{eq37}\end{align}	
where $\delta _\Delta={e^{j2\pi {f_c}[(r_I-r_Q)/c-\tau]}}$, $\delta _E\!=\!{e^{j2\pi {f_c}(r_I-r_Q)/c}}$, and $\sigma ^2$ denotes the noise variance. Then, we choose the maximum SINR of the in-phase symbol component as
\begin{eqnarray}
\begin{aligned}[b]
\Gamma^{x_I}   \buildrel \Delta \over =  \max \{{\Gamma ^{x_I}_1},{\Gamma ^{x_I}_2}\}.
\end{aligned}
\label{eq38}\end{eqnarray} 
Analogous to \eqref{eq37} and \eqref{eq38}, we get the average SINR of the quadrature symbol components as
\begin{align}
	{\left\{{\begin{array}{*{20}{l}}
				{\Gamma _1^{{x_Q}}\! =\! \frac{{{\rm{R}}{{\rm{e}}^2}\{ {\bf{h}}_Q^H{{\bf{w}}_Q}\} }}{{{\rm{I}}{{\rm{m}}^2}\{ {\delta _{ \!-\! \Delta }}{\bf{h}}_I^H{{\bf{w}}_I}\} \! +\! {\rm{R}}{{\rm{e}}^2}\{ {\delta _{ \!- \!}}{\bf{h}}_Q^H{{\bf{n}}_{A,Q}}\} \! +\! {\rm{I}}{{\rm{m}}^2}\{ {\delta _{\! - \!\Delta }}{\bf{h}}_I^H{{\bf{n}}_{A,I}}\}  \!+\! {\sigma ^2}}},}\\
				{\Gamma _2^{{x_Q}}\! =\! \frac{{{\rm{I}}{{\rm{m}}^2}\{ {\bf{h}}_Q^H{{\bf{w}}_Q}\} }}{{{\rm{R}}{{\rm{e}}^2}\{ {\delta _{\! -\! \Delta }}{\bf{h}}_I^H{{\bf{w}}_I}\} \! +\! {\rm{I}}{{\rm{m}}^2}\{ {\delta _{\! -\! }}{\bf{h}}_Q^H{{\bf{n}}_{A,Q}}\} \! +\! {\rm{R}}{{\rm{e}}^2}\{ {\delta _{\! -\! \Delta }}{\bf{h}}_I^H{{\bf{n}}_{A,I}}\} \! +\! {\sigma ^2}}},}
		\end{array}} \right.}
	\label{eq32}\end{align}	
where $\delta_{-\Delta}={e^{-j2\pi {f_c}[(r_I-r_Q)/c-\tau]}}$ and $\delta _{-}={e^{j2\pi {f_c}\tau}}$. We choose the maximum correct recovery of the quadrature symbol component as
\begin{eqnarray}
\begin{aligned}[b]
\Gamma^{x_Q}   \buildrel \Delta \over =  \max \{{\Gamma ^{x_Q}_1},{\Gamma ^{x_Q}_2}\}.
\end{aligned}
\label{eq40}\end{eqnarray}
Then, the total SINR of the received signal can be defined as
\begin{eqnarray}
\begin{aligned}[b]
\Gamma  \buildrel \Delta \over = {\Gamma ^{{x_I}}} + {\Gamma ^{{x_Q}}}.
\end{aligned}
\label{eq41}\end{eqnarray}

We chooes ${\mathcal{S}_E}$ as the potential positions of Eves outside the intersection of the two mainlobes since the passive Eves may be located at arbitrary position $({x_E},{y_E}) \in {\mathcal{S}_E}$. The degree of transmission security can be quantified by adopting the concept of secrecy rate in the information-theoretic sense \cite{Broadcast_Csiszar}. The average secrecy rate, denoted by $R_S$, is the difference between the rate of LU and Eve \cite{Physical_Bloch,Secrecy_Oggier}. Following this definition, we have
\begin{eqnarray}
{R_S} \buildrel \Delta \over = {\left[ {\log (1 + {\Gamma _U}) - \mathop {\max }\limits_{{{\cal S}_E}} \log (1 + {\Gamma _E})} \right]^ + },
\label{eq42}\end{eqnarray}
where ${\Gamma _U}$ and ${\Gamma _E}$ are the average SINR of LU and passive Eves.  

Based on the above discussion, the average SINR of LU is given by 
\begin{eqnarray}
\begin{aligned}[b]
\Gamma_U  = \zeta,
\end{aligned}
\label{eq43}\end{eqnarray}
which is consistent with the received SNR requirements.

The $M$-PSK modulation consists of in-phase and quadrature components of the signal. Using the average SINR in \eqref{eq38} and  \eqref{eq40}, the symbol error probability for any positions equals the probability that either branch has a bit error \cite{Digital_Proakis}, i.e.,
\begin{eqnarray}
{P_s} = 1 - [1 - Q(\sqrt {2{\Gamma ^{{x_I}}}} )][1 - Q(\sqrt {2{\Gamma ^{{x_Q}}}} )],
\label{eq44}\end{eqnarray}
where $Q(z) = \int_z^\infty  {\frac{1}{{\sqrt {2\pi } }}} {e^{ - {{{x^2}}}/{2}}}dx$ is the tail distribution function of the standard normal distribution.

\subsection{Error Analysis}
In practice, there are always some measurement errors in the location of LU or the locations may be outdated for the movement. Then, we analyze the effect of measurement error on performance. Let us focus on latency time caused by the measurement errors of range difference, i.e.,
\begin{eqnarray}
	\begin{aligned}[b]
		\tau  = \hat \tau  + \Delta \tau.
	\end{aligned}
	\label{eq45}\end{eqnarray}
where $\tau$,  $\hat \tau$, and $\Delta \tau$ are the actual, estimated, and measurement error latency time, respectively.

We first analyze the influence of carrier orthogonality on  measurement errors. The received signal at LU under measurement errors can be expressed as
\begin{align}
	{\hat y_U}(t)\! =\!  &\sqrt \zeta  {\sigma _U}{x_I}(t)\cos [2\pi {f_c}(t \! - \! \frac{{{r_{U,I}}}}{c})]\notag \\&
\!+\! \sqrt \zeta  {\sigma _U}{x_Q}(t \!-\! \Delta \tau )\sin [2\pi {f_c}(t \! -\!  \frac{{{r_{U,I}}}}{c} + \Delta \tau )] \! + \! {n_U}.
	\label{eq46}\end{align}	
Then, we obtains the in-phase and quadrature symbol components by down-convert operation and low-pass filter as
\begin{eqnarray}
	\begin{aligned}[b]
		\left\{ {\begin{array}{*{20}{l}}
			\hat	{y_U}(t) \cdot \cos [2\pi {f_c}(t - \frac{{{r_I}}}{c})] \\	\overset{\rm{LP}}\to  {\hat y_{U,I}}(t) = \sqrt \zeta  {\sigma _U} \cdot {x_I}(t) \\ \kern 14pt+ \underbrace {\sqrt \zeta  {\sigma _U}{x_Q}(t - \Delta \tau )\sin (2\pi {f_c}\Delta \tau ) + {n_U}}_{{\rm{Interference}}},\\
			\hat	{y_U}(t) \cdot \sin [2\pi {f_c}(t - \frac{{{r_I}}}{c}+ \Delta \tau)] \\	\overset{\rm{LP}}\to  {\hat y_{U,Q}}(t) = \sqrt \zeta  {\sigma _U} \cdot {x_Q}(t - \Delta \tau )\\ \kern 14pt + \underbrace {\sqrt \zeta  {\sigma _U}{x_I}(t)\sin (2\pi {f_c}\Delta \tau ) + {n_U}}_{{\rm{Interference}}}.
		\end{array}} \right.
	\end{aligned}
	\label{eq47}\end{eqnarray}
It can be easily found that the carrier may be not orthogonal due to the measurement errors, which forms inter-branch interference. When $2\pi {f_c}\Delta \tau  \to  k\pi ,k = 0, \pm 1,...$, the carriers are closer to orthogonal, so there is less inter-branch interference. When $2\pi {f_c}\Delta \tau   \to  k\pi  + \pi /2,k = 0, \pm 1,...$, the carriers gradually lose orthogonality, resulting in greater inter-branch interference. Moreover, the measurement error tolerance is closely related to the transmit carrier frequency. That is, the lower carrier frequency, the larger measurement error tolerance.

If the measurement errors cause the symbols to be misaligned, the LU cannot receive the information correctly. We assume the symbol period is $t_0$. It's easy to draw conclusions that when  $ \left| {\Delta \tau } \right| < {t_0/2}$, the LU is able to recover messages correctly. However, when $ \left| {\Delta \tau } \right| > {t_0/2}$, symbol misalignment causes an error output.

The system can still transmit effectively within a certain error, but its performance will be reduced. In fact, for secure wireless transmission, the robustness goes against the security. The prior location information of the transmitters and the LUs can be obtained in advance. Especially for the scenarios where the transmitters' and LUs' positions are fixed or the trajectory can be predicted, we can use pilot signal, feedback link adjustment, and high precision GPS to minimize the measurement errors as much as possible, so as to improve the system secrecy performance.

\section{Simulation Results}

In this section, we provide numerical simulation results to validate the performance of our proposed D3M scheme. Unless stated otherwise, the simulation parameters used in this section are as follows. The carrier frequency is set to ${f_c} = 1\;{\rm{GHz}}$. For simplicity, all the background thermal noise variance is assumed to be identical, i.e., $10\log (\sigma _U^2)=10\log (\sigma _E^2)=- 100 \;\text{dBm}$, and the distributed transmitters contain a total of $N$ elements, i.e., $N=N_I+N_Q$, $N_I=N_Q$. The antenna spacing of ULA is $d_I=d_Q = c/(2f_c)$ to avoid creating grating lobes. We fix LU's coordinate as $(x_{U},y_{U})=(530\;\rm{m},570\;\rm{m} )$. The conventional single array transmission (SAT) located at the origin presented in \cite{Phased_Zhang, Artificial_Xie} and the conventional null space projection (NSP) method presented in \cite{Robust_Shu} are adopted as performance references, respectively. Following the electromagnetic wave propagation path loss model in the free space \cite{Wireless_Goldsmith}, the path loss factor is determined by
\begin{eqnarray}
\begin{aligned}[b]
{\rm{Lfs}}(\text{dB}) &= - 20{\lg }[\rho (r)]\\&
= 32.5 + 20\lg[{f_c}(\text{MHz})] + 20\lg[r(\text{Km})].
\end{aligned}
\label{eq48}\end{eqnarray}

\begin{figure}[tb]
	\begin{center}
		\includegraphics[width=0.9\columnwidth]{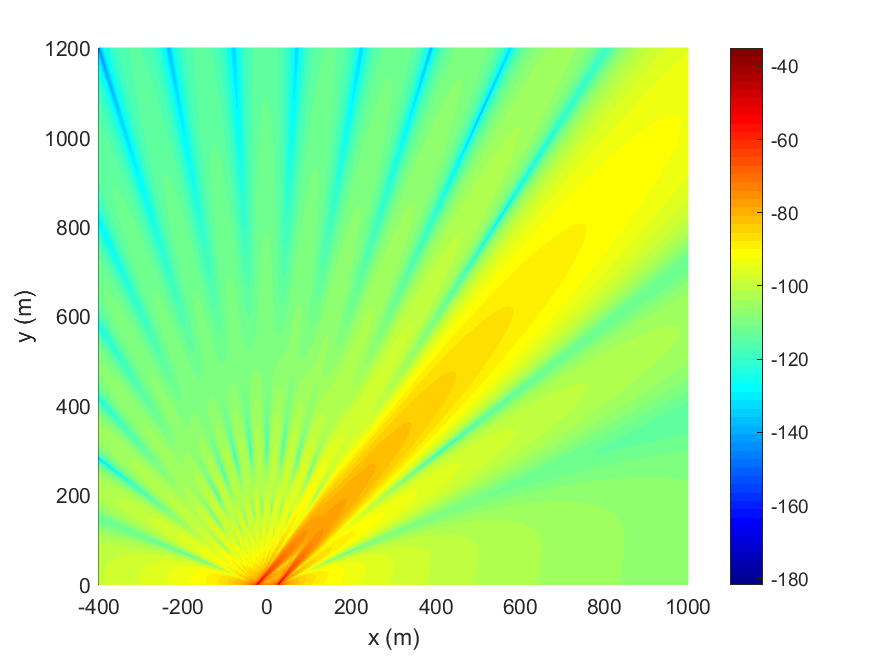}
	\end{center}
	\caption{Message power distribution versus coordinate on the x-y plane, where $N=32$, $D_T=50$ m, and $\zeta=10$ dB.}
	\label{fig4}
\end{figure}
Figure \ref{fig4} depicts the spatial distribution of the received message power, i.e., without AN. As expected, the radiation patterns of two transmit arrays both indicate the desired directions. The received message power synthesized nearby LU is the equivalent of about $-90$ dBm, which is conformed to the received message power requirements. That is, our design of beamforming vector can satisfy the received message power requirements for LU. At the same time, it effectively suppresses the power of received message for Eves along the sidelobes. As the radio wave propagates, the message power gradually weakens. That is, it follows that the closer receiver gets to transmitter the stronger received message power becomes. Therefore, it tends to more message leakage when Eves are close to the transmitter.
\begin{figure}[tb]
	\begin{centering}
		\subfloat[]{\begin{centering}
				\includegraphics[width=0.8\columnwidth]{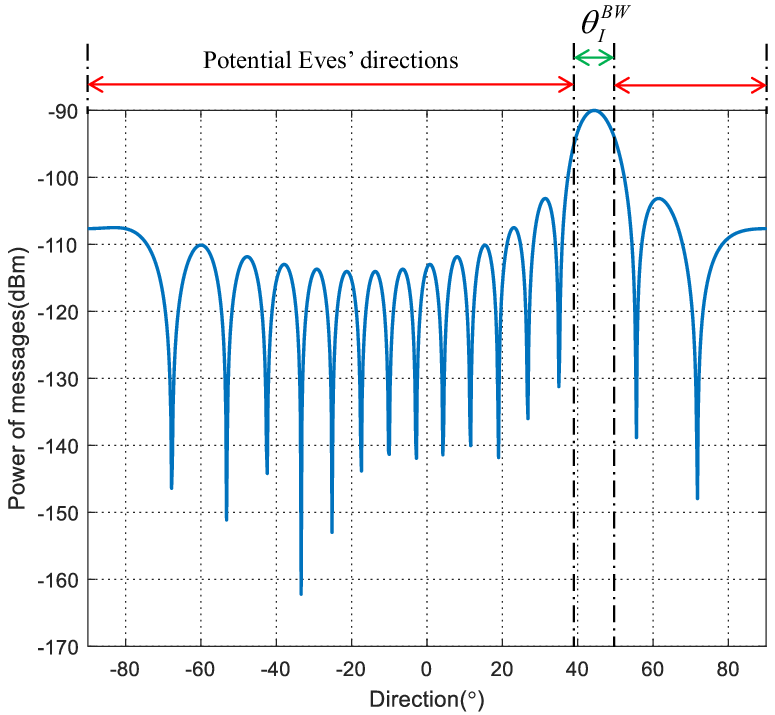}
				\par\end{centering}
		}\\
		\subfloat[]{\begin{centering}
				\includegraphics[width=0.8\columnwidth]{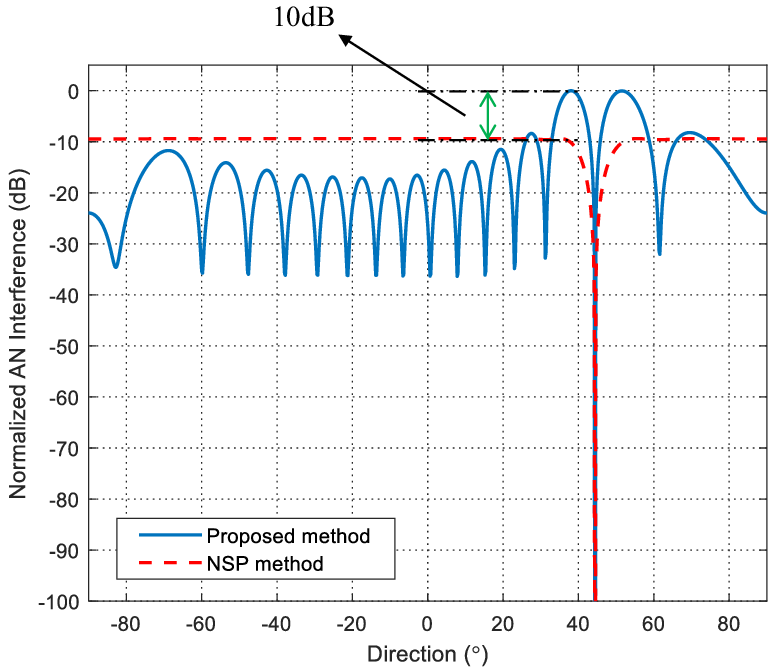}
				\par\end{centering}
		}\\
		\par\end{centering}
	\caption{Power distribution emitted by the transmitter I versus direction dimension (a) message power, (b) AN interference power, where $N=32$ and $\gamma=-20$ dB.}
	\label{fig5}\end{figure}

\begin{figure}[tb]
	\begin{center}
		\includegraphics[width=0.9\columnwidth]{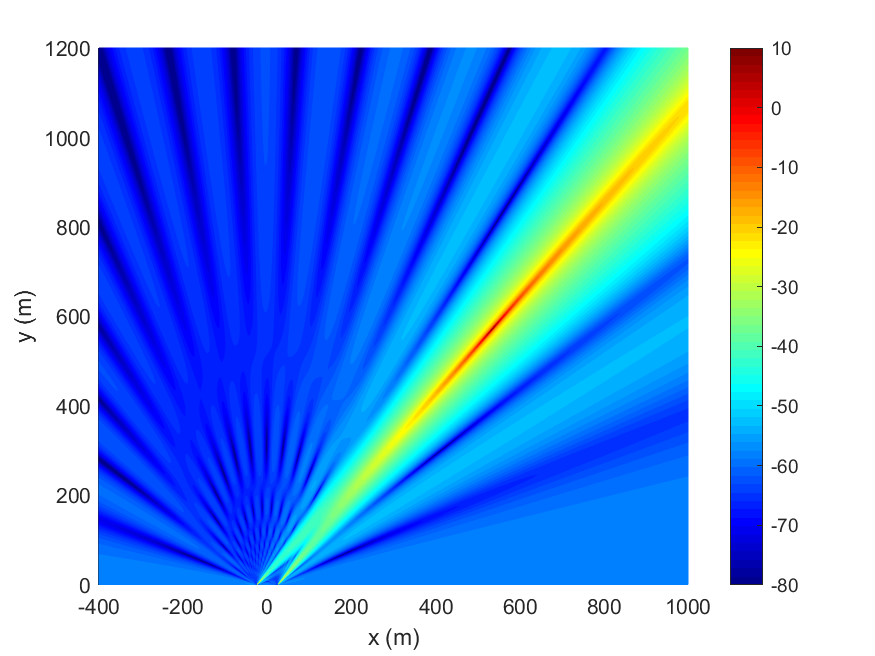}
	\end{center}
	\caption{SINR distribution versus coordinate on the x-y plane, where $N=32$, $D_T=50$ m, $\gamma=-20$ dB, and $\zeta=10$ dB.}
	\label{fig6}
\end{figure}

The message and AN power distributions versus direction dimension are plotted in Fig. \ref{fig5}(a) and \ref{fig5}(b), respectively. From Fig. \ref{fig5}(a), it can be observed that the message power in the desired direction satisfies the required power and the leakage power on the sidelobes is different. In particular, for the proximal desired directions more message power is leaked. In Fig. \ref{fig5}(b), we compare AN interference power of our proposed AN design method with conventional NSP method. For a fair comparison, in the simulations, we set transmit AN power to be equal. A general observation is that a deep null of AN interference power is formed along the direction of LU. This means that the design of AN nulls out the interference to LU. The AN interference power of the NSP method is distributed uniformly in the undesired directions regardless of the message power along side-lobe directions due to the only constraint on the null space of the desired direction. By taking the message power of the sidelobes into account, the proposed AN design method makes full use of the limited transmit AN power to meet the target SINR value in the undesired directions. In general, as Eve moves close to LU, it can readily intercept the confidential messages for more message power leakage at the proximal LU. As seen, the proposed AN design method is more focused on the proximal desired direction, and thus generates more interference for proximal LU, about 10 dB over the NSP method. This makes the potential Eve near LU suffered from more AN interference. Hence, our proposed AN design method enhances the security. Moreover, due to the free space path loss, the closer the Eve gets to transmitter, the more message power gets, so does AN interference power. This causes a uniform and weak average SINR distribution (cf. Fig. \ref{fig6}) outside the mainbeam around the desired location where Eve may exist. Overall, our proposed AN design method is able to guarantee the target SINR in the undesired directions with less transmit AN power consumption, which utilizes AN to provide security with high efficiency gains. Additionally, AN interference power distribution emitted by transmitter Q has the same result, and thus, it will not be shown further. 

We further illustrate the average SINR distribution in Fig. \ref{fig6}. As expected, a SINR peak converges to LU's location on the x-y plane. The value of SINR peak is equal to 10 dB, which satisfies the received SNR requirement for LU. This indicates that high-reliable information-receiving for LU can be guaranteed. As a contrast, the average SINR outside the neighborhood around the LU's location is smooth and much lower than the peak value. This is caused by the weak message leakage power in addition to high AN interference. 
\begin{figure}[tb]
	\begin{center}
		\includegraphics[width=0.9\columnwidth]{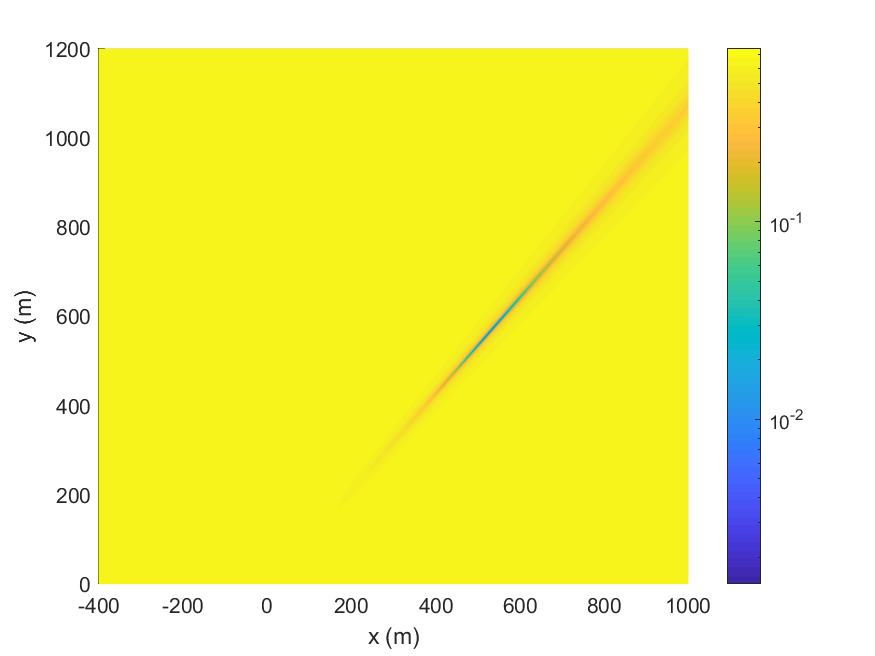}
	\end{center}
	\caption{SER distribution versus coordinate on the x-y plane, where $N=32$, $D_T=50$ m, $\gamma=-20$ dB, and $\zeta=10$ dB.}
	\label{fig7}
\end{figure}

Then, we illustrate the average SER in Fig. \ref{fig7}. We adopt quadrature phase shift keying (QPSK) modulation. Consistent with the SINR results in Fig. \ref{fig6}, the average SER is low only at the position of LU, while our proposed scheme causes a uniform and considerable average SER distribution in other regions. A general observation is that the ultimate aim of spatial (including angle-range-dependent) secure transmission is achieved.

\begin{figure}[tb]
	\begin{center}
		\includegraphics[width=0.8\columnwidth]{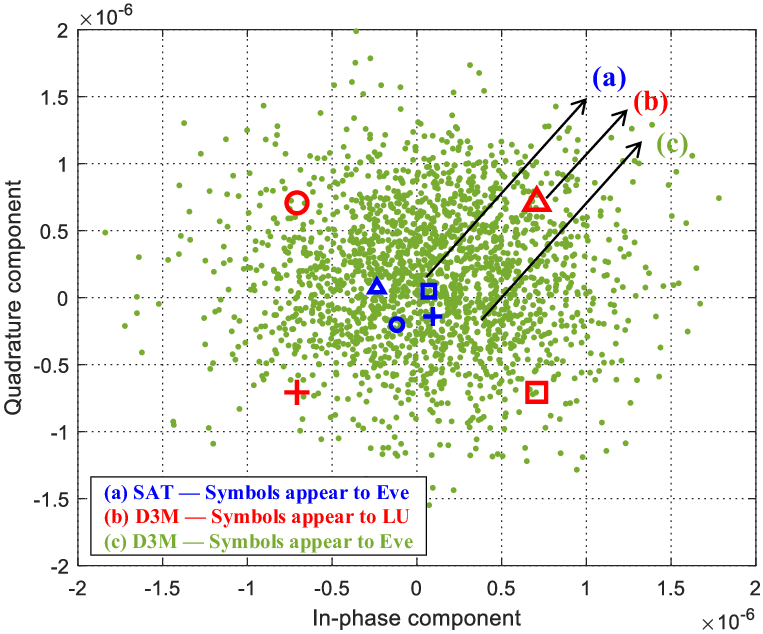}
	\end{center}
	\caption{Constellation diagrams of the noiseless received signals as it appears to Eve and LU.}
	\label{fig8}
\end{figure}

We show the received constellation diagram at LU and typical Eve synthesized by the proposed D3M method to transmit QPSK signals. As is evident in Fig. \ref{fig8}, the constellation produced by the proposed D3M method at LU is standard constellation diagram while, along an undesired transmit locations, the received symbols appear to be random. Also, the constellation produced by conventional SAT method is still crisp enough to demodulate for a sensitive Eve.

\begin{figure}[tb]
	\begin{center}
		\includegraphics[width=0.9\columnwidth]{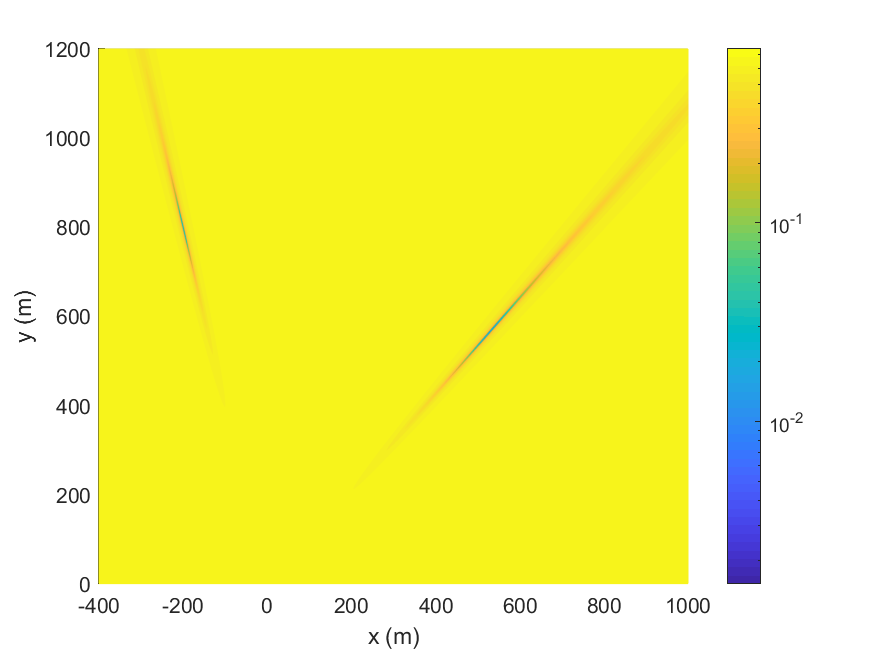}
	\end{center}
	\caption{SER distribution versus coordinate on the x-y plane for multiple users, where $K=2$, $N=32$, $D_T=50$ m, $\gamma=-20$ dB, and $\zeta_1=\zeta_2=10$ dB.}
	\label{fig9}
\end{figure}

We further show the average SER for the case of multiple LUs in Fig. \ref{fig9}. Typically, in the case of two LUs, i.e., $K$=2, who are located at $(x_{U_1},y_{U_1})=(530\;\rm{m},570\;\rm{m} )$, $(x_{U_2},y_{U_2})=(-200\;\rm{m},800\;\rm{m})$, and $\zeta_1=\zeta_2=10$ dB (possibly different). One can observe that a low SER can be formed around LUs' locations whereas a high level of SER is imposed in other locations. This indicates the proposed multi-LU scheme is capable of conveying the multiple message streams to the corresponding LUs, simultaneously, while defending the confidential messages from wiretapping. Therefore, the proposed scheme provides a feasible solution of secure transmission in handling the case of multiuser.

\begin{figure}[tb]
	\begin{center}
		\includegraphics[width=0.8\columnwidth]{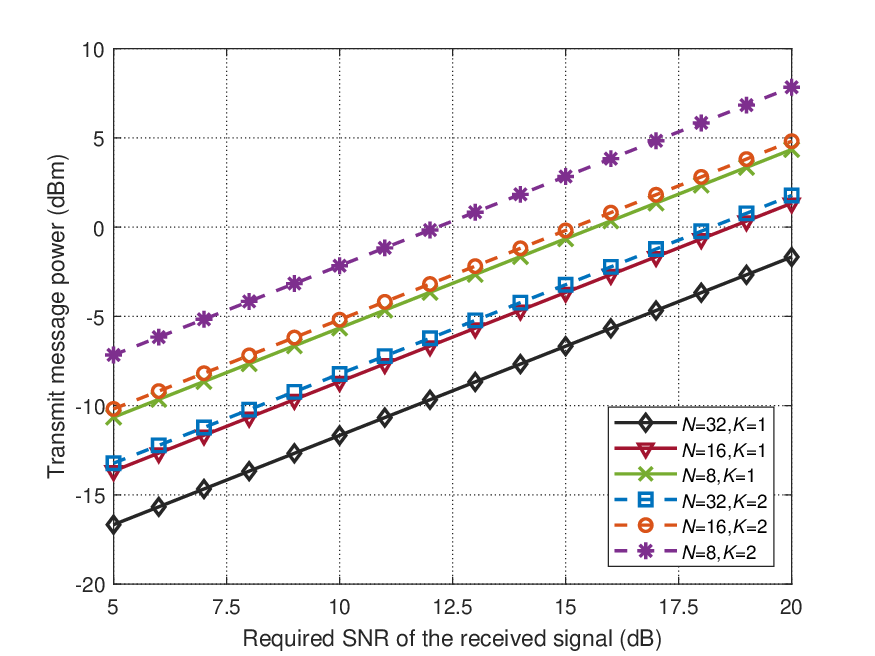}
	\end{center}
	\caption{Minimum transmit message power versus required received SNR, for different numbers of transmit antennas and LUs.}
	\label{fig10}
\end{figure}

\begin{figure}[tb]
	\begin{center}
		\includegraphics[width=0.8\columnwidth]{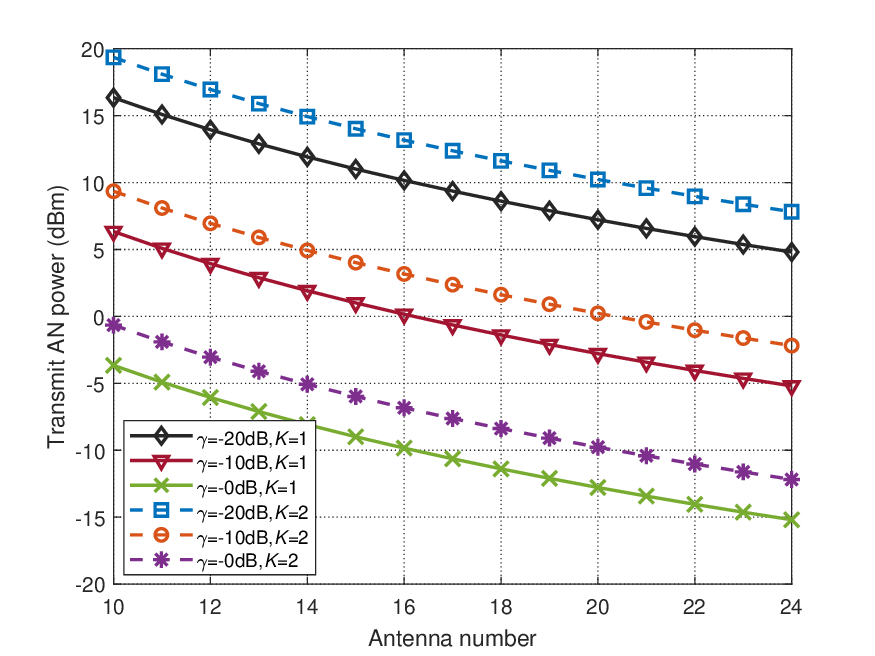}
	\end{center}
	\caption{Minimum transmit AN power versus the number of transmit antennas, for different target SINR and numbers of LUs.}
	\label{fig11}
\end{figure}

Figures \ref{fig10} and \ref{fig11} illustrate how the number of transmit antennas and LUs influences the power consumption, including the transmit message power and transmit AN power. Unless otherwise stated, the consumed power refers to total power of two distributed transmitters. The message power consumption versus the required received SNR of LUs for different number of transmit antennas and LUs is shown in Fig. \ref{fig10}. It can be seen that, a larger value of required received SNR at LUs indicates that it has to consume more transmit message power to meet the quality of transmission for LUs. Besides, it is noticed that, larger number of transmit antennas yields less transmit message power consumption, and more LUs need more transmit message power. Our proposed design of the beamforming vector is to provide a prescribed quality communication assurance for LUs with the minimum transmit message power. That is, given a prescribed minimum received SNR value, the minimum transmit message power is determined, which is able to minimize the potential of message leakage to Eve. The transmit AN power consumption versus the number of transmit antennas, for different target SINR in the undesired locations and numbers of LUs, is shown in Fig. \ref{fig11}. Similarly, lower target SINR for the undesired directions, less transmit antennas, and more LUs are required for more transmit AN power consumption. 
\begin{figure}[tb]
	\begin{center}
		\includegraphics[width=0.8\columnwidth]{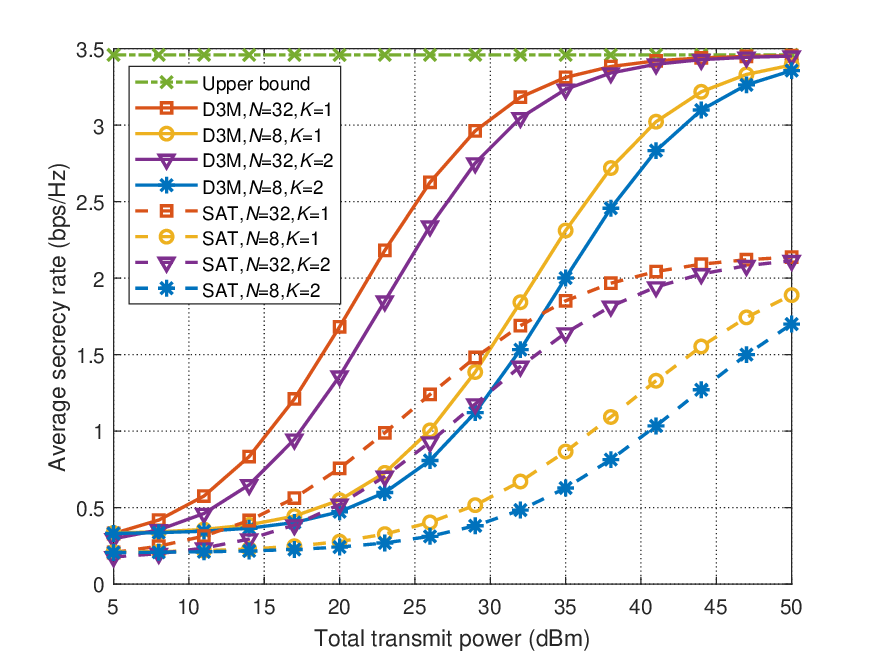}
	\end{center}
	\caption{Average secrecy rate versus the total tansmit power, for different number of transmit antennas, where $D_T=50$ m, $\gamma=-20$ dB, and $\zeta=10$ dB.}
	\label{fig12}
\end{figure}

Figure \ref{fig12} presents the average secrecy rate of our proposed scheme versus the total transmit power, for different numbers of transmit antennas and LUs  as well as those of SAT scheme. Moreover, we also present the theoretic upper bound of average secrecy rate; that is, the achievable rate without Eve, which is given by ${\log_2}(1+\zeta)$. To evaluate the security performance for multiuser scenario, the average secrecy rate is adopted as secrecy-sum-rate \cite{Robust_Shu,Secrecy_Oggier}, which can be defined as ${R_S} \buildrel \Delta \over =\frac{1}{K} \sum\limits_{k = 1}^K {{{\left[ {\log \left( {1 + {\Gamma _{U,k}}} \right) - \mathop {\max }\limits_{{\mathcal{S}_E}} \log \left( {1 + {\Gamma _{E,k}}} \right)} \right]}^ + }}$. From this figure, we see that larger total transmit power yields higher average secrecy rate. Obviously, the average secrecy rate is relatively low in small total transmit power regime due to the fact that in such a case, insufficient available transmit power can be consumed to produce AN to reduce achievable rate of Eve. However, in high transmit power regime, the average secrecy rate of our proposed scheme tends to saturate and its value converges to the theoretical upper bound. The reason is that more transmit power can be consumed to produce the AN, which makes the available rate of Eve close to zero. What's more, as is well-known in array signal processing, adding more antennas to the transmit array enhances the array's capability in the degree of spatial freedom thereby enabling a thinned main-lobe width to improve the average secrecy rate. On the other hand, it is evident that the average secrecy rate for the SAT scheme is much lower than that of the proposed D3M scheme. This is because the beampattern of phased-array transmission is only angle-dependent. We consider that Eve is located anywhere, including identical direction as the LU. The SAT scheme fails to provide satisfactory secrecy performance when LU and Eve are located along the same direction. Therefore, it is more probable to enhance the security for our proposed D3M scheme compared to the SAT scheme in practice.

\section{Conclusion}
In this paper, we proposed an AN-aided D3M secure transmission scheme to enhance PLS in wireless communication systems against passive eavesdropping. Based on the phased-array transmission structure, the transmit $M$-PSK constellation was decomposed into in-phase and quadrature components, and then modulated them into mutually orthogonal branches, each of which was transmitted by one transmitter. Specifically, the beamforming vector of the transmitters was individually designed to ensure that the signal branches can be synchronic and accurate transmission for the LU with the minimum transmit message power. We employed an iteration algorithm which optimized the beamforming vectors of transmitters successively. A closed form solution for each step showed that the proposed algorithm can be efficiently implemented. Next, the AN projection matrix was imposed to further scramble the received signal at Eve. Moreover, our proposed scheme was further extended to the cases of multi-LU. Finally, extensive numerical simulations demonstrated the significant performance advantages. Our proposed AN-aided D3M scheme can achieve ``point'' secure transformation, whose security overcomes ``line'' secure transformation of the conventional phased array transmission. Due to the limitation of LoS transformation channel and high-precision position information, our proposed D3M scheme can be applied to the near future static or quasi-static LoS transmission scenarios with high security requirements.

\ifCLASSOPTIONcaptionsoff
  \newpage
\fi

\bibliographystyle{IEEEtran}
\bibliography{IEEEabrv,REF}

\end{document}